\documentclass[11pt]{article}

\usepackage{amsmath,amssymb}
\usepackage{graphicx}

\usepackage{physics}
\usepackage{centernot}
\usepackage{enumitem}
\usepackage{amsthm}

\usepackage{tabularx,booktabs,adjustbox,array}
\usepackage{longtable}
\newcolumntype{Y}{>{\raggedright\arraybackslash}X}

\usepackage{adjustbox}
\usepackage{siunitx}
\ExplSyntaxOn
\msg_redirect_name:nnn { siunitx } { physics-pkg } { none }
\ExplSyntaxOff
\usepackage{tikz}
\usetikzlibrary{arrows.meta,positioning,shapes.geometric,fit,calc}
\usepackage{pgfplots}
\pgfplotsset{compat=1.18}

\usepackage{soul}

\usepackage[T1]{fontenc}
\usepackage{lmodern}
\usepackage{microtype}
\usepackage{xurl}
\usepackage{cite}

\newcommand{\gev}{\giga\electronvolt}
\newcommand{\fsl}[1]{{\centernot{#1}}}
\newcommand\numberthis{\addtocounter{equation}{1}\tag{\theequation}}

\newcounter{bla}

\newcommand{\hi}{\texttt{HyperIso}}
\newcommand{\spi}{\texttt{SuperIso}}
\newcommand{\mty}{\texttt{MARTY}}

\definecolor{myg}{RGB}{56, 140, 70}

\newcommand{\releasecommit}{%
  \href{https://github.com/Hyperiso/Hyperiso/commit/09d26ac6e71d1528922097a896d15693589f636b}%
       {\texttt{09d26ac6e71d}}}
\newcommand{\releasedate}{20 July 2026}

\newcommand{\softwaredoi}{%
  \href{https://doi.org/10.5281/zenodo.21447072}%
       {\nolinkurl{10.5281/zenodo.21447072}}}

\newcommand{\releaseversion}{v1.0.3}
\newcommand{\packageversion}{1.0.3}

\usepackage[most]{tcolorbox}
\tcbuselibrary{listings}

\usepackage[hidelinks]{hyperref}
\hypersetup{
  pdftitle  = {HyperIso: A general BSM calculator for flavour observables},
  pdfauthor = {T. Reymermier, N. Fardeau, F. Mahmoudi}
}
\usepackage{xurl}

\lstdefinelanguage{json}{
  basicstyle=\ttfamily\small,
  string=[s]{"}{"},
  stringstyle=\color{green!40!black},
  comment=[l]{//},
  commentstyle=\color{gray},
  morecomment=[s]{/*}{*/},
  literate=
   *{0}{{{\color{blue!60!black}0}}}{1}
    {1}{{{\color{blue!60!black}1}}}{1}
    {2}{{{\color{blue!60!black}2}}}{1}
    {3}{{{\color{blue!60!black}3}}}{1}
    {4}{{{\color{blue!60!black}4}}}{1}
    {5}{{{\color{blue!60!black}5}}}{1}
    {6}{{{\color{blue!60!black}6}}}{1}
    {7}{{{\color{blue!60!black}7}}}{1}
    {8}{{{\color{blue!60!black}8}}}{1}
    {9}{{{\color{blue!60!black}9}}}{1}
}

\lstdefinelanguage{yaml}{
  basicstyle=\ttfamily\small,
  sensitive=true,
  comment=[l]{\#},
  commentstyle=\color{gray},
  morestring=[b]",
  morestring=[b]',
  stringstyle=\color{green!40!black},
  morekeywords={true,false,null,yes,no,on,off},
  keywordstyle=\color{blue!60!black}\bfseries
}

\lstdefinestyle{arxivcode}{
  basicstyle=\ttfamily\small,
  numbers=left,
  numberstyle=\tiny\color{gray},
  numbersep=8pt,
  breaklines=true,
  breakatwhitespace=false,
  showstringspaces=false,
  keepspaces=true,
  columns=fullflexible,
  upquote=true,
  tabsize=2
}

\tcbset{
  arxivcodebox/.style={
    listing only,
    listing engine=listings,
    colback=black!2,
    colframe=black!20,
    boxrule=0.3pt,
    arc=1mm,
    left=2mm,
    right=2mm,
    top=1mm,
    bottom=1mm
  }
}

\newtcblisting{cppcode}{
  arxivcodebox,
  listing options={style=arxivcode,language=C++}
}

\newtcblisting{pycode}{
  arxivcodebox,
  listing options={style=arxivcode,language=Python}
}

\newtcblisting{terminal}{
  arxivcodebox,
  listing options={style=arxivcode,language=bash}
}

\newtcblisting{jsoncode}{
  arxivcodebox,
  listing options={style=arxivcode,language=json}
}

\newtcblisting{yamlcode}{
  arxivcodebox,
  listing options={style=arxivcode,language=yaml}
}

\newtcblisting{lhacode}{
  arxivcodebox,
  listing options={style=arxivcode}
}

\begin{document}

\begin{center}
\Large\bf\boldmath
\vspace*{2.0cm}\hi: A general BSM calculator for flavour observables
\unboldmath
\end{center}

\vspace{0.8cm}
\begin{center}
N. Fardeau\textsuperscript{a}, F. Mahmoudi\textsuperscript{a,b,c},
T. Reymermier\textsuperscript{a,}\footnote{Corresponding author: \tt theo.reymermier@cern.ch\\
\hspace*{0.48cm} URL: \tt http://hyperiso.in2p3.fr}\\[1cm]
{\sl \textsuperscript{a} Université Lyon 1, CNRS, IP2I, UMR 5822, Villeurbanne, France\\[.2cm]
\textsuperscript{b} Theoretical Physics Department, CERN, CH-1211 Geneva 23, Switzerland\\[.2cm]
\textsuperscript{c} Institut Universitaire de France (IUF), 75005 Paris, France
}
\end{center}

\vspace{0.9cm}
\begin{abstract}
\noindent We present \hi, a new standalone program for the evaluation of flavour physics observables across a wide variety of Standard Model (SM) and Beyond-the-Standard-Model (BSM) scenarios. The code builds on the physics heritage of \spi, but is implemented as a new modular software package with a modern C++ core and multiple user interfaces. In addition to native implementations for established scenarios such as the SM, the general Two-Higgs-Doublet Model (THDM), and supersymmetric models, \hi\ interfaces with the \mty\ framework to compute automatically BSM Wilson coefficients at leading order. This design makes it possible to study user-defined BSM models while keeping the observable calculation, uncertainty propagation and statistical interpretation in a common backend. The software provides C++, Python, command-line and graphical interfaces. It supports the evaluation of key observables such as the branching ratios and, where available, angular observables, for radiative, leptonic, and semileptonic \(B\) decays, as well as for the relevant kaon and \(D\)-meson decays, together with the muon anomalous magnetic moment \((g-2)\). The inputs can be provided through JSON/YAML configuration files and Les Houches Accord files, either supplied by the user or generated by external spectrum calculators. \hi\ therefore provides a flexible and extensible tool for flavour studies, phenomenological scans and the validation of BSM scenarios.
\end{abstract}
\vfill
\pagebreak
\section{Introduction}

The study of flavour physics observables plays an important role in testing the Standard Model (SM) and exploring potential extensions, such as Supersymmetry (SUSY), Two-Higgs-Doublet Model (THDM), and other Beyond-the-Standard-Model (BSM) scenarios. These observables, including branching ratios of rare decays and the muon anomalous magnetic moment \((g-2)\), provide sensitive probes of new physics effects. With increasing experimental precision at experiments such as LHCb, Belle II and BaBar, theoretical tools capable of handling complex scenarios are essential for accurately constraining BSM scenarios by predicting observables in both standard and non-standard models and making statistical analysis.

Existing software tools such as \spi~\cite{Mahmoudi:2007vz,Mahmoudi:2008tp,Mahmoudi:2009zz,Neshatpour:2021nbn} have made significant strides in enabling such analyses. However, their scope is often limited to specific models, and their extensibility to novel theoretical models or new effective theories remains constrained. Furthermore, user interfaces in these tools are typically rigid, restricting accessibility for diverse users, from phenomenologists to experimentalists.

To address these limitations, we present \hi~\cite{HyperIsoSoftware101,Reymermier:2026tym}, a comprehensive software package for evaluating flavour-physics observables in the Standard Model and in user-defined BSM scenarios, and for performing statistical analyses. By combining a C++ core with the \mty~framework~\cite{Uhlrich:2020ltd,Uhlrich:2020aaj,Uhlrich:2021ded} for generic leading-order Wilson-coefficient calculations, \hi\ provides an extensible and high-performance calculation environment. C++, Python, command-line and graphical interfaces make the same backend accessible to phenomenologists, experimentalists and software developers. \hi\ accepts Les Houches Accord (LHA) \cite{Skands:2003cj, Allanach:2008qq, Mahmoudi:2010iz}, family inputs and can be integrated with spectrum calculators and parameter-scan workflows.

This paper describes the design, implementation and public interfaces of \hi, together with its validation, performance and reproducibility workflow. For legacy formulae, we refer to the original \spi~documentation~\cite{Mahmoudi:2008tp}. Numerical comparisons have been carried out to validate both the common calculations with \spi~(Wilson coefficients, observables and fits) and the \mty{} generic implementation for Wilson coefficients.

The remainder of this paper is organised as follows.
Section~\ref{sec:design} presents the design goals and architectural principles.
Sections~\ref{sec:architecture} and~\ref{sec:program_description} describe the
software architecture, runtime data flow and main program components.
Section~\ref{sec:input_structure} introduces the input conventions, while
Section~\ref{sec:user_guide} presents the C++, Python, command-line and
graphical interfaces. Section~\ref{sec:theory} summarises the theoretical
framework and statistical methods. Sections~\ref{sec:validation}
and~\ref{sec:performance} report validation and performance results.
Finally, Section~\ref{sec:software_availability} documents software
availability and the frozen reproducibility suite, and we summarise our conclusions in Section~\ref{sec:conclusion}.

\section{Design goals and principles}\label{sec:design}

\begin{figure}[!ht]
    \centering
    \includegraphics[width=1\linewidth]{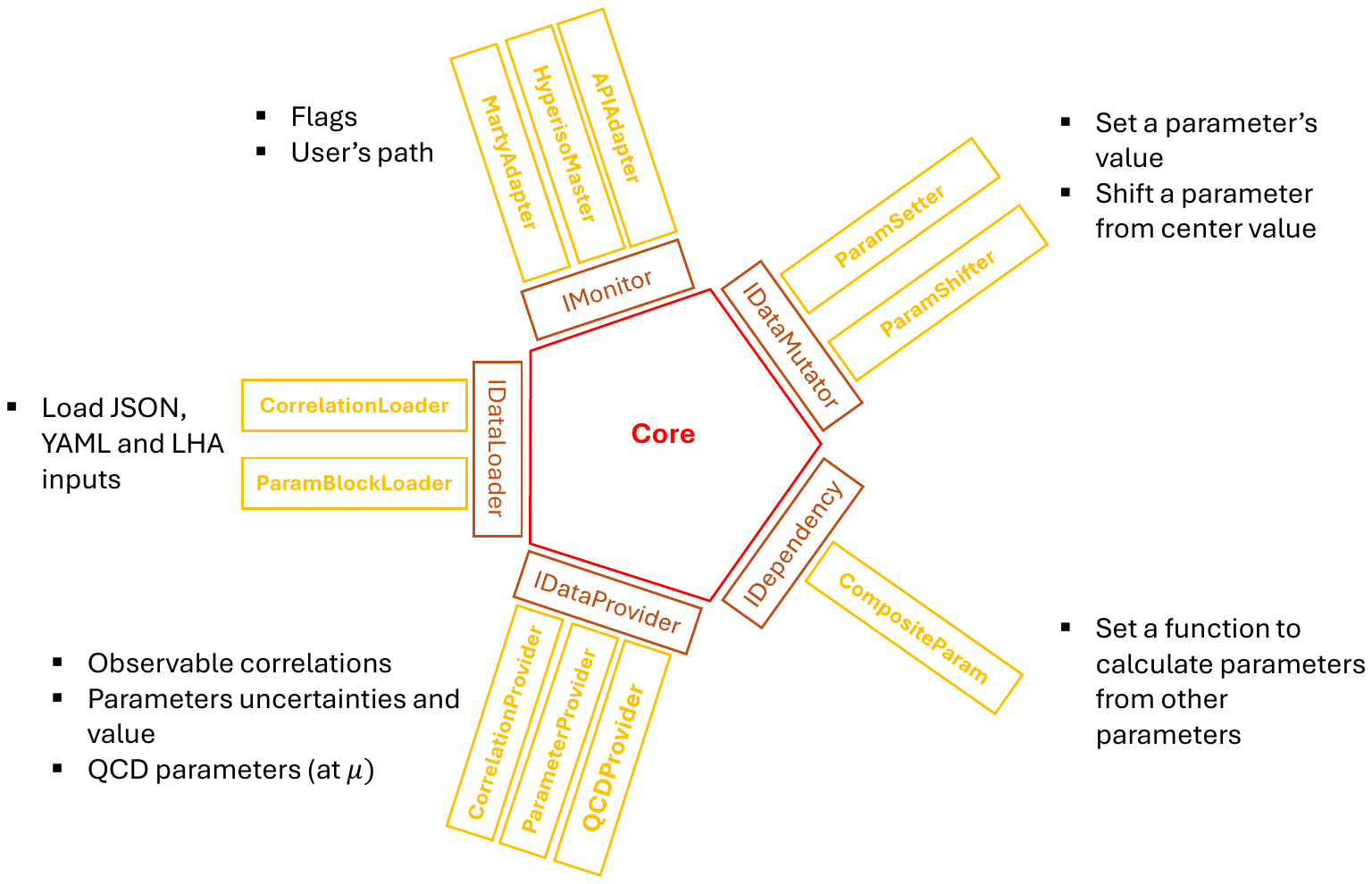}
    \caption{Hexagonal architecture of the \hi\ Core module.}
    \label{fig:core_architecture}
\end{figure}

\hi~follows a hexagonal, or ports-and-adapters, architecture~\cite{Cockburn2005Hexagonal}. The application logic is organised into modules dedicated to runtime data, Wilson coefficients, observable calculations and statistics. Each module exposes ports that define the services required or provided by the domain, while adapters connect those ports to file formats, external programs and public interfaces. This separation limits coupling to infrastructure and allows the domain logic to be tested independently. The architecture of the Core module is illustrated in Fig.~\ref{fig:core_architecture}.

This architecture reduces coupling between modules and makes the physics, input, interface and numerical layers easier to test independently.

\section{Software architecture}\label{sec:architecture}

\begin{figure}[!t]
    \centering
    \includegraphics[width=1\linewidth]{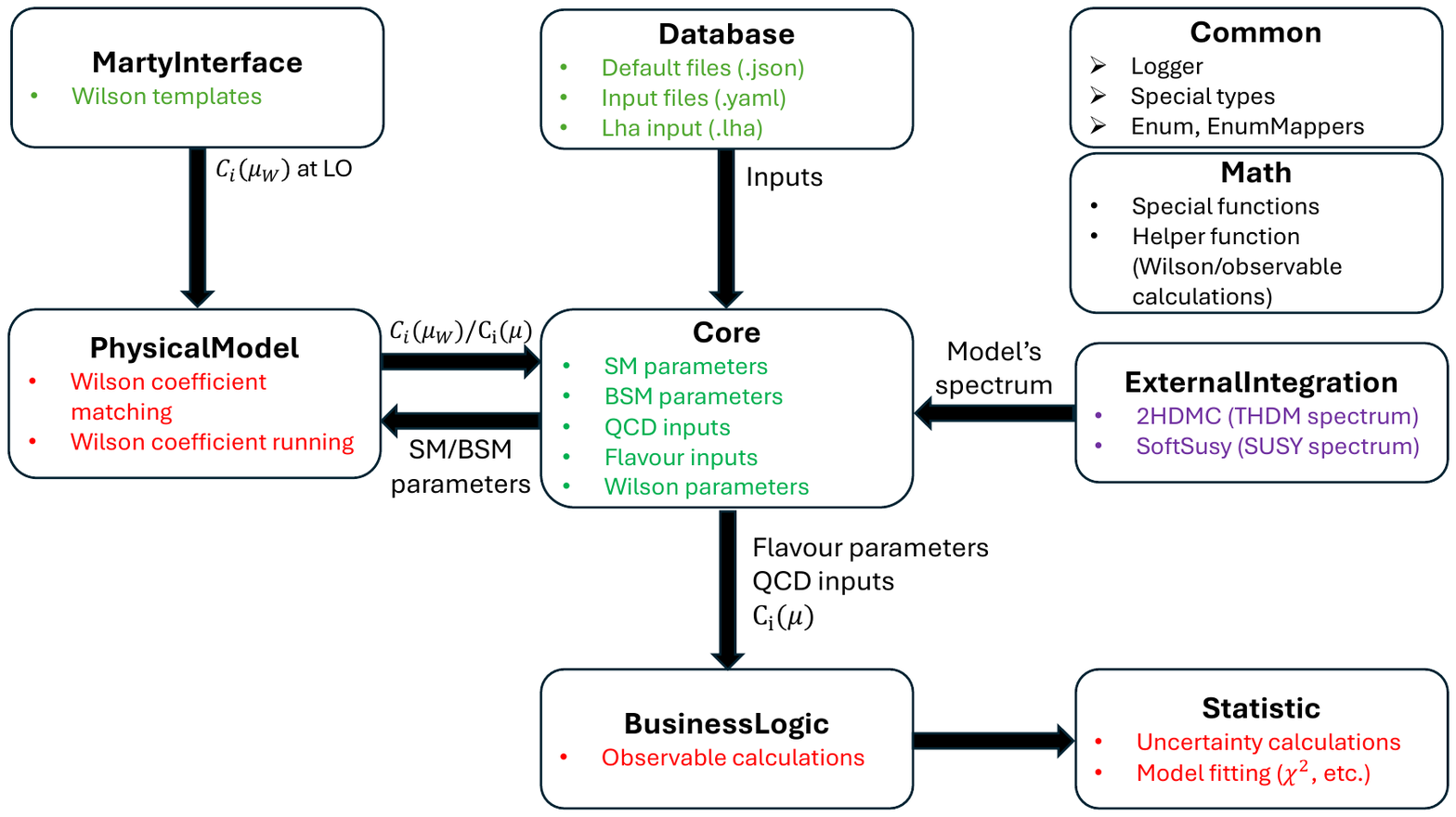}
    \caption{Data flow in the \hi\ software.}
    \label{fig:hyperiso_data_flow}
\end{figure}

A high-level description of the \hi{} architecture and data flow is shown in Fig.~\ref{fig:hyperiso_data_flow}. The distributed JSON files provide the default numerical database used by the code, including SM inputs, QCD parameters, flavour parameters, nuisance distributions, observable definitions and correlations. These defaults are intended to define a reproducible reference setup and should not be edited directly by users.

User-level modifications are provided through YAML override files. This mechanism is useful when a parameter value, uncertainty or distribution has to be updated to match a given experimental input, global fit or phenomenological convention. The override layer is deliberately separated from the distributed defaults so that scans can be reproduced by archiving only the user-provided YAML files and the LHA input.

Model-specific parameters are supplied through LHA, SLHA or FLHA files \cite{Skands:2003cj,Allanach:2008qq,Mahmoudi:2010iz}. These files are loaded after the JSON defaults and YAML overrides; entries found in the LHA-family input therefore have the highest priority for overlapping quantities.

\section{Program description}\label{sec:program_description}

\hi~is a modular software tool designed for the computation of Wilson coefficients, flavour observables and statistical analyses. Its architecture is built around a C++ core and applies the SOLID object-oriented design principles~\cite{Martin2003Agile, martin2017clean}, with several user-facing layers built on top of the same internal calculation pipeline.

The program supports four complementary interfaces:
\begin{itemize}
    \item \textbf{C++ interface.} The native interface gives direct access to the core and is intended for developers or users who want to write custom \texttt{main} programs, implement new observables, or integrate \hi~into a larger compiled workflow.
    \item \textbf{Python interface.} The C++ core is exposed to Python through \texttt{pybind11}. This interface is intended for scripted scans, notebooks, automated comparisons and integration with Python-based phenomenological workflows.
    \item \textbf{Command-line interface.} A command-line layer is provided for simple calculations and quick checks, without writing a dedicated C++ or Python driver.
    \item \textbf{Graphical interface.} The GUI is based on Dash and targets interactive use. It gives access to Wilson-coefficient calculations, observable predictions, uncertainty estimates, Monte-Carlo uncertainty propagation, maximum-likelihood fits, confidence contours and parameter-space exploration.
\end{itemize}

All interfaces rely on the same initialisation pattern: a \texttt{HyperisoMaster} object is created and initialised with a runtime configuration and an LHA-family input file. The object must remain alive while the dependent Wilson, observable and statistical interfaces are in use. Packaged assets and user path overrides are resolved through the active runtime path provider installed by \texttt{HyperisoMaster}; installed Python wheels therefore do not depend on paths from the machine or container on which they were compiled. This guarantees that all frontends use the same parameter-loading rules, Wilson-coefficient pipeline and observable calculators.

\section{Input structure and data conventions}
\label{sec:input_structure}

The numerical inputs of \hi~are stored in typed parameter blocks.  Whenever possible, block names and integer identifiers follow the FLHA convention; additional non-FLHA blocks use the same block--identifier logic with project-defined IDs.  The full catalogue of default blocks and custom identifiers is given in \ref{app:block_structure}.  At runtime, the same block database is shared by the C++, Python, CLI and graphical interfaces.

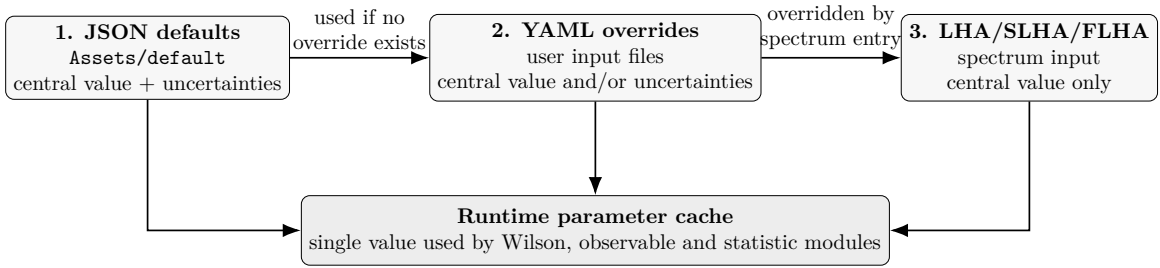
\begin{figure}[!ht]
\centering
\resizebox{\textwidth}{!}{%
\begin{tikzpicture}[
  font=\small,
  node distance=2.1cm,
  source/.style={draw, rounded corners, align=center, minimum width=3.5cm, minimum height=1.25cm, fill=black!3},
  cache/.style={draw, rounded corners, align=center, minimum width=4.0cm, minimum height=1.15cm, fill=black!7},
  arrow/.style={-{Latex[length=3mm]}, thick}
]
\node[source] (defaults) {\textbf{1. JSON defaults}\\\texttt{Assets/default}\\central value + uncertainties};
\node[source, right=2.3cm of defaults] (yaml) {\textbf{2. YAML overrides}\\user input files\\central value and/or uncertainties};
\node[source, right=2.3cm of yaml] (lha) {\textbf{3. LHA/SLHA/FLHA}\\spectrum input\\central value only};
\node[cache, below=1.55cm of yaml] (runtime) {\textbf{Runtime parameter cache}\\single value used by Wilson, observable and statistic modules};
\draw[arrow] (defaults) -- node[above, align=center] {used if no\\override exists} (yaml);
\draw[arrow] (yaml) -- node[above, align=center] {overridden by\\spectrum entry} (lha);
\draw[arrow] (defaults.south) |- (runtime.west);
\draw[arrow] (yaml.south) -- (runtime.north);
\draw[arrow] (lha.south) |- (runtime.east);
\end{tikzpicture}%
}
\caption{Input precedence in \hi.  JSON files provide the distributed reference database.  YAML files document user-level changes, including uncertainty metadata.  LHA-family files provide the final model-dependent central values and override matching entries in the runtime cache.}
\label{fig:input_precedence}
\end{figure}

The input system is deliberately layered.  Distributed JSON files define the reference database used by a clean installation.  User YAML files may override selected entries, including their central value and uncertainty metadata.  Finally, the LHA/SLHA/FLHA spectrum file passed to \texttt{HyperisoMaster::init(...)} is used as the model-specific numerical input.  This last layer has the highest priority for overlapping entries, but LHA-family files provide only the numerical value read from the spectrum block; they do not carry the separate statistical and systematic uncertainty fields used by the internal database.  The resulting precedence rule is summarised in Fig.~\ref{fig:input_precedence}.

The JSON layer is version controlled with the code rather than edited for each scan.  A typical entry stores the central value and the uncertainty fields used by the statistical layer:
\begin{jsoncode}
{
  "SMINPUTS": {
    "6": {
      "central_value": 172.69,
      "stat_error": 0.30,
      "syst_error": 0.40
    }
  }
}
\end{jsoncode}
A YAML override has the same logical structure but is supplied by the user.  It is therefore a compact way of recording an alternative input convention without modifying the distributed defaults:
\begin{yamlcode}
SMINPUTS:
  6:
    central_value: 171.5
    stat_error: 1.0
    syst_error: 0.0
\end{yamlcode}
The corresponding value in a spectrum file follows the LHA-family syntax and overrides the runtime central value when the same block and identifier are present:
\begin{lhacode}
Block SMINPUTS
    6   1.71500000E+02   # m_t
\end{lhacode}
In this case, the LHA value becomes the active top-quark mass in the cache.  The uncertainty metadata remains controlled by the JSON/YAML database and by the statistical configuration.  YAML files are therefore best suited for non-LHA quantities, nuisance parameters, uncertainty models or explicit studies of alternative conventions.  LHA, SLHA and FLHA files are best suited for model-specific spectra and Wilson-coefficient inputs.

For BSM models, the LHA parser can be extended before initialisation through the \texttt{pre\_init} interface of \texttt{HyperisoMaster}.  This is useful when a model introduces spectrum blocks that are not part of the standard LHA/SLHA/FLHA prototypes.  The relevant call is
\begin{cppcode}
HyperisoMaster hyp;
hyp.pre_init_add_block("SPECIAL_BLOCK",
                       2,   // itemCount
                       1,   // valueIdx
                      -1,   // scaleIdx
                      -1,   // rgIdx
                      -1,   // binIdx
                       false); // globalScale
hyp.init("model_input.flha", config);
\end{cppcode}
The same method is exposed in Python as \texttt{pre\_init\_add\_block}. Its arguments describe how a line of the custom block is mapped to an internal parameter entry; they are summarised in Table~\ref{tab:preinit_block_args}. The column indices are zero-based, consistent with the internal \texttt{Prototype} representation. The method should be called before \texttt{init()}, so that the first spectrum parsing already uses the custom prototype.

\begin{table}[!ht]
\centering
\small
\renewcommand{\arraystretch}{1.15}
\begin{tabularx}{\textwidth}{@{}p{0.22\textwidth}p{0.16\textwidth}X@{}}
\toprule
\textbf{Argument} & \textbf{Default} & \textbf{Meaning} \\
\midrule
\texttt{blockName} & -- & Case-insensitive name of the custom LHA block. \\
\texttt{itemCount} & \texttt{2} & Total number of tokens expected on each data line of the block. \\
\texttt{valueIdx} & \texttt{1} & Zero-based column index containing the numerical value to store. \\
\texttt{scaleIdx} & \texttt{-1} & Optional zero-based column index for a local scale; \texttt{-1} means that no per-entry scale is present. \\
\texttt{rgIdx} & \texttt{-1} & Optional zero-based column index for a renormalisation or scheme identifier. \\
\texttt{binIdx} & \texttt{-1} & Optional zero-based column index for binned entries; the parser uses it to encode bin information in the internal identifier. \\
\texttt{globalScale} & \texttt{false} & Set to \texttt{true} when the block uses a global \texttt{Q=} scale in the block header. \\
\bottomrule
\end{tabularx}
\caption{Arguments of the custom-block registration routine.  They define the LHA prototype used by the parser for BSM-specific blocks.}
\label{tab:preinit_block_args}
\end{table}

Paths to external tools and writable runtime directories can also be registered before initialisation. The method \texttt{pre\_init\_set\_marty\_path} selects an existing \mty~installation, while \texttt{pre\_init\_set\_softsusy\_path} selects a \texttt{SOFTSUSY} executable or installation directory.  The lower-level \texttt{pre\_init\_set\_paths} routine overrides selected internal paths, such as the MARTY and spectrum-cache directories.  The active LHA path is intentionally excluded from this routine because it is supplied explicitly to \texttt{init()} or to the LHA-switching routine.

\subsection{FLHA conventions for observables}

We expand the FLHA convention \cite{Mahmoudi:2010iz} in order to account for the observables exposed by the public \hi~interfaces. For binned observables, six additional integer fields encode the optional lower and upper bounds of the $q^2$ bin in GeV, with three integer fields assigned to each edge. This avoids floating-point identifiers while keeping sub-GeV bin definitions reproducible. We also extend the observable-type field beyond the values defined in the original FLHA convention. The public identifiers generated by the \releaseversion~mapper are gathered in Table~\ref{tab:obs_ids}; internal or unreleased observable identifiers are intentionally omitted.

\begin{table}[!ht]
    \centering
    \small
    \renewcommand{\arraystretch}{1.12}
    \begin{tabularx}{\textwidth}{@{}p{0.19\textwidth}X@{}}
        \toprule
        \textbf{Type ID} & \textbf{Observable type generated by the public \releaseversion~ mapper} \\
        \midrule
        \texttt{15} & Untagged branching fraction. \\
        \texttt{50} & Zero crossing of $A_{FB}$. \\
        \texttt{51}, \texttt{52}, \texttt{53} &
        Lepton-side, hadron-side and mixed forward--backward asymmetries
        $A_{FB}^{\ell}$, $A_{FB}^{h}$ and $A_{FB}^{\ell h}$. \\
        \texttt{91ij} &
        Lepton-universality or charged-current ratio $R_M^{ij}$; the current
        mapper emits the concrete values \texttt{9121} and \texttt{9132}. \\
        \texttt{92ij} &
        Polarisation, where $i=1$ denotes a fermion (lepton), $i=2$ a
        longitudinally polarised vector and $i=3$ a transversely polarised
        vector; $j$ is the one-based position of the particle of interest in
        the ordered daughter list. \\
        \texttt{931} & Longitudinal polarisation fraction $F_L$. \\
        \texttt{932} & Transverse fraction $F_T$. \\
        \texttt{933} & Angular coefficient $\alpha_K$. \\
        \texttt{934} & Flat term $F_H$. \\
        \texttt{941i}, \texttt{942}, \texttt{943i} &
        Transverse asymmetries $A_T^{(i)}$, $A_{\rm Im}$ and $H_T^{(i)}$. \\
        \texttt{951i}, \texttt{952i}, \texttt{953i} &
        Angular observables $P_i$, $P'_i$ and $S_i$; multi-index variants are
        used for $S_{1c}$, $S_{2s}$, $S_{2c}$ and $S_{6c}$. \\
        \bottomrule
    \end{tabularx}
    \caption{Additional FLHA observable-type identifiers exposed by the public
    HyperIso \releaseversion~ mapper. For CP-asymmetric angular counterparts, the mapper
    uses the negative of the corresponding positive type code. Polarisation
    observables follow the project-defined \texttt{92ij} convention.}
    \label{tab:obs_ids}
\end{table}
\newpage
For example, the $\tau$ in $B\to D^{(*)}\tau\nu$ is the second daughter and
therefore uses type \texttt{9212}, while the longitudinally polarised $D^*$ in
$B\to D^*\tau\nu$ is the first daughter and uses type \texttt{9221}.

\section{User guide and interfaces}\label{sec:user_guide}

The public interfaces of \hi~are designed as lightweight frontends to the same calculation backend. They differ in the degree of interactivity and in the amount of control exposed to the user, but they all rely on the same initialisation step: a \texttt{HyperisoMaster} object is created, an LHA input file is loaded, and a \texttt{HyperisoConfig} object fixes the model and runtime flags. This common entry point is important for reproducibility, since Wilson coefficients, observables and statistical routines then use the same parameter cache and the same precedence rules for default JSON values, YAML overrides and LHA inputs.

\subsection{Overview of the user interfaces}

The four user-facing layers are summarised in Table~\ref{tab:user_interfaces}. The command-line interface is intended for stable built-in workflows, whereas the Python and C++ interfaces expose the programmable API. In particular, custom decays, custom observables and dynamic Wilson groups are treated as API examples rather than as command-line primitives.

\begin{table}[!ht]
\centering
\begin{adjustbox}{max width=\textwidth}
\begin{tabularx}{\textwidth}{@{}p{0.13\textwidth}p{0.35\textwidth}p{0.44\textwidth}@{}}
\toprule
Interface & Typical use & Representative distributed examples \\
\midrule
C++ API & Native compiled workflows, integration in external C++ scans, development of custom observables or Wilson groups. & {\footnotesize\begin{tabular}[t]{@{}l@{}}\texttt{examples\_cpp/Core}\\\texttt{examples\_cpp/Wilson}\\\texttt{examples\_cpp/Observable}\\\texttt{examples\_cpp/Statistic}\end{tabular}} \\
Python API & Interactive and scripted phenomenological analyses, notebooks, scans, post-processing and quick validation. & {\footnotesize\begin{tabular}[t]{@{}l@{}}\texttt{examples\_python/Core}\\\texttt{examples\_python/Wilson}\\\texttt{examples\_python/Observable}\\\texttt{examples\_python/Statistic}\end{tabular}} \\
CLI & Reproducible built-in summaries of Wilson coefficients, observables and statistical uncertainties without writing a driver program. & {\footnotesize\begin{tabular}[t]{@{}l@{}}\texttt{hyperiso-ui wilson summary}\\\texttt{hyperiso-ui observable summary}\\\texttt{hyperiso-ui statistic summary}\end{tabular}} \\
Dash GUI & Interactive exploration of model inputs, Wilson coefficients, observables, uncertainties, best-fit routines and parameter scans. & Graphical frontend built on top of the same Python/C++ backend. \\
\bottomrule
\end{tabularx}
\end{adjustbox}
\caption{Main user interfaces of \hi~and their intended use.}
\label{tab:user_interfaces}
\end{table}

\subsection{Common initialisation}

The initialisation step is deliberately minimal. In Python, the high-level controller is imported from \texttt{pyhyperiso.Core}; model and flag enums are imported from \texttt{pyhyperiso.Common}. The following example initialises a Standard Model run from the reference LHA file distributed with the code:

\begin{pycode}
from pyhyperiso.Common import Model
from pyhyperiso.Core import HyperisoConfig, HyperisoMaster

config = HyperisoConfig(model=Model.SM)
hyp = HyperisoMaster()
hyp.init(lha_file="lha/si_input.flha", config=config)
\end{pycode}

The corresponding C++ entry point follows the same structure:

\begin{cppcode}
#include "HyperisoMaster.h"
#include "Include.h"

HyperisoConfig config;
config.model = Model::SM;

HyperisoMaster hyp;
hyp.init("lha/si_input.flha", config);
\end{cppcode}

The same configuration object also stores runtime flags and model-backend options:
\begin{itemize}[leftmargin=1.5em]
  \item spectrum and input flags, such as \texttt{IS\_LHA\_SPECTRUM}, \texttt{HAS\_WILSON\_INPUT} and \texttt{HAS\_TH\_OBSERVABLE\_INPUT};
  \item MARTY backend flags, such as \texttt{HYP\_AS\_SM\_MARTY};
  \item the model choice, selected with \texttt{Model.MARTY} in Python or \texttt{Model::MARTY} in C++;
  \item the MARTY model name and file path, supplied through \texttt{mty\_model\_name} and \texttt{mty\_model\_path}.
\end{itemize}
External MARTY paths and extra non-FLHA blocks can be declared before initialisation with the \texttt{pre\_init} routines described above. The \texttt{HyperisoMaster} instance must remain alive for as long as the interfaces created from its runtime providers are used.

\subsection{Python interface}

The Python interface exposes the main C++ functionality through \texttt{pybind11} bindings and thin high-level wrappers. The most common modules are \texttt{pyhyperiso.Core}, \texttt{pyhyperiso.Wilson}, \texttt{pyhyperiso.Observable} and \texttt{pyhyperiso.Statistic}. They respectively cover initialisation and parameter access, Wilson coefficients, physical observables, and uncertainty propagation or fits.  The following compact example combines the Wilson and observable interfaces:

\begin{pycode}
from pyhyperiso.Common import (
    Model, WGroup, WCoeff, QCDOrder, ContributionType, ScaleType, WilsonBasis,
    Observables, Decays
)
from pyhyperiso.Core import HyperisoConfig, HyperisoMaster
from pyhyperiso.Wilson import WilsonBuildConfig, WilsonInterface, WilsonRequest
from pyhyperiso.Observable import ObservableInterface

hyp = HyperisoMaster()
hyp.init(lha_file="lha/si_input.flha", config=HyperisoConfig(model=Model.SM))

W = WilsonInterface()
W.build(WilsonBuildConfig(
    groups={WGroup.B}, matching_scale=81.0,
    hadronic_scale=2.0, order=QCDOrder.LO))
req = WilsonRequest(
    group=WGroup.B, coefficient=WCoeff.C9, order=QCDOrder.LO,
    contribution=ContributionType.TOTAL, scale_type=ScaleType.HADRONIC,
    wilson_basis=WilsonBasis.STANDARD)
print(W.get_FR(req))

O = ObservableInterface()
O.add_observable(Observables.BR_B_XS_GAMMA, QCDOrder.NNLO)
O.add_observables_from_decay(Decays.B__l_l, QCDOrder.NNLO)
print(O.compute_observable(Observables.BR_BS_MUMU))
\end{pycode}

For statistical studies, an \texttt{ObservableInterface} is first filled with the observables entering the fit. It is then passed to a \texttt{StatisticInterface} together with a \texttt{StatisticConfig}. The main public methods illustrated in the examples are \texttt{compute\_uncertainties()}, \texttt{compute\_MLE(...)} and \texttt{compute\_confidence\_contour(...)}, respectively used for Monte-Carlo uncertainty estimates, best fits and two-dimensional confidence regions. Experimental data can be restricted either by experiment name or by exact measurement identifiers. An exact identifier is represented by an \texttt{ExperimentObs} object containing an experiment name and a \texttt{BinnedObservableId}; \texttt{select\_experiment\_observables(...)} configures this selection and \texttt{selected\_experiment\_observables()} exposes the retained entries. This prevents unrelated measurements of the same theoretical observable from entering a fit.

\subsection{C++ interface}

The C++ examples are built as independent executables against an installed \hi~package. A minimal build uses CMake and the exported targets of the library:

\begin{terminal}
mkdir -p build
cd build
cmake ..
cmake --build .
\end{terminal}

The example CMake project calls \texttt{find\_package(Hyperiso REQUIRED)} and links to the relevant component targets, such as \texttt{Hyperiso::CoreLib}, \texttt{Hyperiso::PhysicalModelLib}, \texttt{Hyperiso::BusinessLogicLib} and, when available, \texttt{Hyperiso::StatisticLib}. The following C++ fragment shows the same Wilson-coefficient workflow as above:

\begin{cppcode}
#include <iostream>
#include <unordered_set>

#include "HyperisoMaster.h"
#include "Include.h"
#include "WilsonInterface.h"

HyperisoConfig config;
config.model = Model::SM;

HyperisoMaster hyp;
hyp.init("lha/si_input.flha", config);

WilsonBuildConfig wc_config(
    std::unordered_set<WGroup>{WGroup::B, WGroup::BScalar},
    81.0, 2.0, QCDOrder::LO
);

WilsonInterface W;
W.build(wc_config);
std::cout << W.getFR(WGroup::B, WCoef::C9,
                     QCDOrder::LO, ContributionType::TOTAL) << std::endl;
\end{cppcode}

\noindent Observable calculations are handled by \texttt{ObservableInterface}, using \texttt{add\_observable(...)} for individual observables, \texttt{add\_observables(...)} for complete decays and \texttt{BinnedObservableId} for binned quantities. The statistical API mirrors the Python workflow through \texttt{StatisticConfig}, \texttt{StatisticInterface}, \texttt{compute\_uncertainties()}, \texttt{compute\_MLE(...)} and \texttt{compute\_confidence\_contour(...)}. The more development-oriented examples, stored in \texttt{dev\_examples}, illustrate dynamic identifiers, lambda-defined Wilson groups, custom decays and diagnostic utilities.

\subsection{Command-line interface}

The command-line interface is provided by the \texttt{hyperiso-ui} executable. Since it is aimed at stable built-in workflows, it focuses on three families of summaries: Wilson coefficients, observables and statistical dependency or uncertainty estimates. The executable is enabled with the dedicated CLI build flag,

\begin{terminal}
cmake -S Hyperiso/Hyperiso/core -B build -DBUILD_WITH_CLI=ON
cmake --build build --target hyperiso-ui --parallel
\end{terminal}

and accepts the common options \texttt{--model SM|THDM|MSSM|MARTY}, \texttt{--lha <file>} and \texttt{--order LO|NLO|NNLO}. For example, a compact Wilson-coefficient summary can be obtained with

\begin{terminal}
./build/hyperiso-ui wilson summary \
  --model SM \
  --lha reproducibility/inputs/sm_reference.flha \
  --groups BCoefficients \
  --coeffs C7,C9,C10 \
  --qmatch 81 \
  --q 4.8 \
  --order NNLO
\end{terminal}

while representative observable and statistics runs are

\begin{terminal}
./build/hyperiso-ui observable summary \
  --model SM \
  --lha reproducibility/inputs/sm_reference.flha \
  --observables BR_Bs__mu_mu,BR_B__Xs_gamma \
  --order NNLO

./build/hyperiso-ui statistic summary \
  --model SM \
  --lha reproducibility/inputs/sm_reference.flha \
  --observables BR_Bs__mu_mu,BR_B__Xs_gamma \
  --chi2 --uncertainties --progress --draws 200 --seed 123456
\end{terminal}

Binned observables are passed through \texttt{--bins} using the syntax \texttt{OBS:min:max}, for example \texttt{F\_L\_B0\_\_K*0\_mu\_mu:1.1:6.0}. Help is available at the top level and for each command family through \texttt{hyperiso-ui --help}, \texttt{hyperiso-ui wilson --help}, \texttt{hyperiso-ui observable --help} and \texttt{hyperiso-ui statistic --help}.

\subsection{Graphical interface}

The Dash graphical interface provides a no-code route to the same workflows. It is mainly intended for exploratory use: the user selects the model configuration and LHA input, computes Wilson coefficients or observables, propagates uncertainties, launches MLE/Monte-Carlo fits with a $\chi^2$ backend, and scans selected parameters. Representative screenshots of the Wilson, observable, uncertainty and scan panels are distributed in the source-tree examples and online API documentation, together with the corresponding CLI or Python commands used to reproduce the same calculations.

\section{Theoretical framework}\label{sec:theory}

Most of the theoretical calculations used in \hi~are detailed in \spi~documentation. In this section, we merely recall the general framework in which the observables are calculated, and we detail the new statistical engine that is featured by \hi.\smallbreak 
The computation of flavour observables is done within the weak effective field theory in which the heavy ($W,Z,h,t$ and any BSM particle with mass greater than or equal to the $W$ boson) degrees of freedom are integrated out to yield an effective Hamiltonian. This Hamiltonian takes the form of a sum of effective operators with effective numerical coefficients, called \textit{Wilson coefficients} in front of them. For the $b\to s\gamma$ decays the relevant effective Hamiltonian reads \cite{Chetyrkin:1997gb}
\begin{equation}
    \mathcal{H}(b\to sX) = -\frac{4 G_F}{\sqrt{2}} V_{tb}V_{ts}^*\sum_{i=1}^{8} C_i(\mu)\mathcal{O}_i(\mu)+\mathrm{h.c.},
\end{equation}
while for the $b\to s\ell\ell$ decays it includes semileptonic vectorial $\mathcal{O}_{9,10}$ and scalar $\mathcal{O}_{Q_1,Q_2}$ operators \cite{Yan:2000dc}. \hi~also implements other effective Hamiltonians for charged-current $P\to\ell\nu$ decays \cite{Sakaki:2013bfa}, where $P$ stands for any pseudoscalar meson, as well as $\Delta F=2$ neutral-meson mixing operators \cite{Aebischer:2020dsw} and $s\to d\nu\nu$ decays \cite{Brod:2010hi}. \smallbreak
The Wilson coefficients are expanded perturbatively in the strong coupling strength,
\begin{equation}
    C_i(\mu) = \sum_{j=0}^\infty \left(\frac{\alpha_s(\mu)}{4\pi}\right)^j C_{i}^{(j)}(\mu),
\end{equation}
where in practice this expansion is known up to $j=2$.
The Wilson coefficients are first calculated at a scale $\mu_W\sim M_W$ by matching the full and the effective theories, and ran down to the hadronic scale $\mu_b\sim m_b$ (for $B$ meson decays) using the known renormalisation group equations (RGEs) \cite{Aebischer:2020dsw,Czakon:2006ss,Bobeth:2013tba}. 

In \hi~there are two ways to compute the Wilson coefficients. The SM, 2-HDM and MSSM contributions are natively implemented up to NNLO as legacy from \spi, and their calculation can be triggered by using the corresponding \texttt{Model} flag when initialising \texttt{HyperisoMaster}. In any model, one can also use \mty~to compute the LO BSM contribution to the Wilson coefficients. In this case, the SM contribution can either be computed at LO using \mty~as well, or at NNLO using the analytical expressions from \spi~(recommended). This is triggered by using the \texttt{Model::MARTY} flag and the choice of the SM values is set by the flag \texttt{HYP\_AS\_SM\_MARTY}. 

In the generic-model workflow, the BSM contribution is generated directly
from diagrams containing at least one non-SM particle. The total coefficient
is then constructed as the sum of this genuine BSM contribution and the SM
contribution selected through \texttt{HYP\_AS\_SM\_MARTY}. This avoids defining
the BSM contribution through the subtraction of two independently evaluated
SM predictions. For each coefficient, the interface first probes the
tree-level amplitude. If at least one tree-level contribution is present, the
matching stops at that order; otherwise the coefficient is evaluated at one
loop. Empty tree-level amplitudes are detected before Wilson-basis projection.

The semileptonic coefficient \(C_9\) retains a dedicated separation of
photon- and non-photon-penguin contributions required by the projection
conventions used in the interface.
\subsection{Observables}

\hi~natively implements the most constraining flavour observables, with particular emphasis on the $B$-meson decay channels measured at current flavour facilities. It also provides extension mechanisms for user-defined observables and effective coefficients. \smallbreak
The base structure encapsulating all observable calculation mechanics is called \texttt{DecayParent}. This structure stores useful information such as the scale and QCD order at which the observable needs to be calculated, the Wilson coefficient groups it depends on, and it has internal methods to trigger the calculation of the appropriate Wilson coefficients. The \texttt{DecayParent} class also stores a polymorphic configuration structure, which can then be specified in derived decays to pass specific flags or attributes to the calculator. Each physical decay is then described by a derived class from \texttt{DecayParent} and implements the actual calculation of the observables related to this decay. Table~\ref{tab:decays} gives a compact overview of the main native decay families and representative observables, while a more detailed summary can be found in \ref{app:observable_catalogue}.\smallbreak
Whenever a \texttt{DecayParent} subclass is asked to evaluate one of its observables, it first triggers the calculation of the needed Wilson coefficients at the right low-energy scale, then retrieves all the needed parameters from the core memory, and finally performs all the calculations using these inputs. Whenever possible, \hi{} uses the GNU Scientific Library (GSL) \cite{gsl} in order to benefit from its optimised algorithms. In practice, most special functions, except for the highest-order polylogarithms, and all numerical integrals are evaluated using GSL.\smallbreak
Notable changes from \spi~\texttt{v5.0} include:
\begin{itemize}
    \item Generalised $P\to\ell\nu$ decay calculator (see \ref{app:plnu}).
    \item Generalised form-factor and QCD factorisation calculators for all $B\to V$ and $B\to P$ neutral-current transitions ($V$ denotes any vector meson such as $K^*$ or $\phi$). This allows us to remove duplicated code by centralising common sub-calculations among similar decays.
\end{itemize}

\begin{table}[!ht]
\centering
\small
\renewcommand{\arraystretch}{1.15}
\begin{tabularx}{\textwidth}{@{}p{0.20\textwidth}X X@{}}
\toprule
\textbf{Sector} & \textbf{Native decay families} & \textbf{Representative outputs} \\
\midrule
Rare and radiative $B$ decays &
$B\to X_s\gamma$, $B\to X_s\ell\ell$, $B\to K^{(*)}\ell\ell$, $B_s\to\phi\ell\ell$, $\Lambda_b\to\Lambda\ell\ell$ &
Branching fractions, differential rates, forward--backward asymmetries, angular coefficients and optimised observables. \\
\addlinespace
Leptonic and charged-current decays &
$B_{s,d}\to\ell\ell$, $B_u\to\ell\nu$, $B\to D^{(*)}\ell\nu$ &
Leptonic branching fractions, $R(D^{(*)})$, polarisation and angular observables. \\
\addlinespace
Kaon, charm and mixing observables &
$K\to\pi\nu\bar\nu$, $K\to\pi\ell\ell$, $K_{L,S}\to\mu\mu$, $K\to\ell\nu$, $D_{(s)}\to\ell\nu$, neutral-meson mixing &
Rare-kaon branching fractions, leptonic ratios, $D_{(s)}$ leptonic modes, mass differences and CP phases. \\
\addlinespace
Extensions &
Custom decay and observable registration layer &
User-defined observables attached to native or custom decay families. \\
\bottomrule
\end{tabularx}
\caption{Compact overview of the main observable sectors natively covered by \hi. A more detailed decay-by-decay catalogue is given in \ref{app:observable_catalogue}; the exact canonical identifiers are distributed with the observable mapper and can be queried through the public interfaces.}
\label{tab:decays}
\end{table}

\subsection{Statistics}
One of the major improvements of \hi~ is found in its completely overhauled statistical engine. Most HEP software packages rely on Gaussian approximation and $\chi^2$ statistics for their prediction and fitting routines. However, an increasing amount of experimental data is available with non-Gaussian uncertainties, ranging from simple asymmetric split-Gaussian distributions $y={\bar y}^{\,+\sigma_+}_{\,-\sigma_-}$ to full likelihood profiles. We expect that taking a conservative Gaussian estimate with the largest standard deviation leads to a decrease in the fit power by overestimating some uncertainties.

The major difficulty that arises when one wants to take into account non-Gaussian distributions is dealing with correlations properly. Indeed, correlation matrices given by experimental fits are calculated under the Gaussian approximation, and only describe joint multivariate-normal distributions. If one wants to consider different marginal distributions for the individual components, one has to also change the way correlations are described. In \hi, this is achieved using \textit{copulas}, which are mathematical tools which allow us to decouple the marginal distributions from the correlation structure of a random vector.

Mathematically speaking, a \textit{copula} $C$ is a joint cumulative distribution function (CDF) of a $d$-dimensional random vector on $[0,1]^d$ with uniform $\mathcal{U}(0,1)$ marginals. These objects can be used to describe any multivariate probability distribution, as stated by Sklar's theorem~\cite{Sklar1959}: \smallbreak
\textbf{Theorem} (Sklar) Let $X=(X_1,\dots,X_d)\in \mathbb R^d$ be a random vector with joint CDF $F(x)=\mathbb{P}(X_1\leq x_1,\dots,X_d\leq x_d)$ and marginal CDFs $F_i(x_i)=\mathbb{P}(X_i\leq x_i)$. There exists a copula $C:[0,1]^d\to[0,1]$ such that:
\begin{equation}
    F(x)=C(F_1(x_1),\dots,F_d(x_d)).
\end{equation}
If the marginal CDFs are continuous, then the copula $C$ is unique. Assuming the existence of the derivatives, the joint probability density function (PDF) $h(x)$ of $X$ reads:
\begin{equation}
    h(x)=\partial_1\dots\partial_dC(F_1(x_1),\dots,F_d(x_d))\prod_{i=1}^d F_i'(x_i).
\end{equation}
\smallbreak
Using copulas, we will then be able to choose independently the marginal distributions of our input variables and their correlation structure. In practice, as experimental input correlation is often given in the form of a correlation matrix with an underlying Gaussian assumption, we will use a Gaussian copula to model the correlation. The Gaussian copula has the following form, given a $d$-dimensional positive definite correlation matrix $R$:
\begin{equation}
    C(u)=\Phi_d(\Phi^{-1}(u_1),\dots,\Phi^{-1}(u_d)),
\end{equation}
where $\Phi_d$ is the joint CDF of a multivariate normal distribution with mean vector $0$ and covariance matrix $R$, and $\Phi^{-1}$ is the quantile function of the standard normal distribution.

\subsubsection{Inputs}
In order to perform its statistical evaluations, \hi~ takes the following inputs:
\begin{enumerate}
    \item The experimental measurements of the implemented observables, separated in a vector of expected values $\mathcal{O}_{\rm exp.}$ and an experimental covariance matrix $\Sigma_\mathcal{O}$ (Gaussian assumption),
    \item The central values $\eta_0$ and covariance matrix $\Sigma_\eta$ of the nuisance parameters if they are Gaussian, else the marginal distribution of each nuisance parameter which can be either Gaussian, split-Gaussian or flat. 
\end{enumerate}
The statistical distributions of both the nuisance parameters and the experimental observables are modelled using copulas taking into account the different marginal distributions and the global covariance of the parameters. This allows us to relax the Gaussian assumption while still modelling the correlations accurately. The central values and standard deviations of the experimental observables are stored in the \texttt{observables.json} asset file which should be overridden by adding entries in the \texttt{observables.yaml} file. Similarly, the correlations among experimental observable values are stored in the \texttt{observables\_corr}, and the correlation among input parameters in the \texttt{parameters\_corr} files.
\subsubsection{Observable prediction}
Given a set of model/fit parameters $p$, \hi\ uses a Monte-Carlo sampling of the nuisance parameter distributions to evaluate the distribution of model predictions. \hi\ then computes the skewness of each marginal distribution corresponding to the various observables, and decides whether it can be assumed to be Gaussian or not. If the skewness is below a user-defined threshold, set by the \texttt{skew\_abs\_threshold} parameter in the \texttt{StatisticConfig}, a Gaussian approximation is made and the prediction for a given observable is: 
\begin{equation}
    \mathcal{O}_{\rm th}=\mu\pm s,
\end{equation}
where $\mu$ is the population mean of the distribution and $s$ is the unbiased population standard deviation. If the skewness is above the threshold, then the result is returned in the form of a split-Gaussian distribution
\begin{equation}
    \mathcal{O}_{\rm th}=m^{+\sigma_+}_{-\sigma_-},
\end{equation}
with mode $m$, and asymmetric standard deviations $(\sigma_-,\sigma_+)$.\smallbreak
This calculation is launched by the \texttt{compute\_uncertainties()} method of the \texttt{StatisticInterface} once the latter has been fully initialised with the desired observables. The number of Monte-Carlo samples is configured through \texttt{StatisticConfig::MC\_draws}. Empirically, a value of \num{500} to \num{1000} samples has been found to yield stable estimates for the observable uncertainties, while as few as 50 samples provide a reliable central value and a rough estimate of the uncertainty.
\subsubsection{Model fitting}
\begin{figure}[ht]
    \centering
    \includegraphics[width=0.75\linewidth]{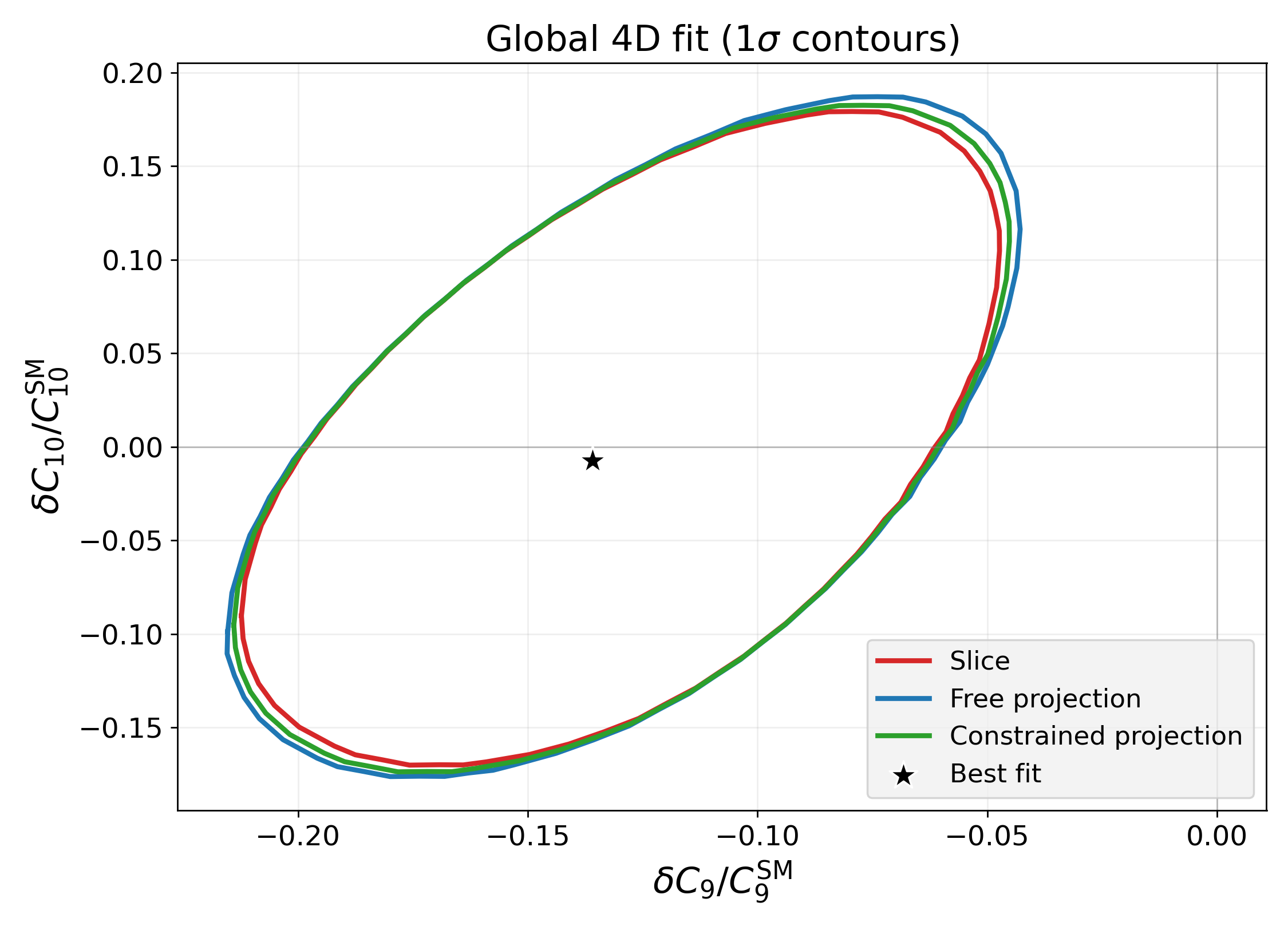}
    \caption{Effect of the three projection methods on the $(\delta C_9,\delta C_{10})$ $1\sigma$ confidence contour from a four-dimensional fit including the scalar operators $C_{Q_1},C_{Q_2}$ to CMS measurements of the angular distributions of the $B^0\to {K^*}^0\mu^+\mu^-$ decay.}
    \label{fig:contours_proj}
\end{figure}

Another use case of \hi~ is to perform a maximum-likelihood fit for some set of parameters $p$ (which can be model parameters, form factors, Wilson coefficients\dots), given a set of nuisance parameters $\eta$ with known statistical distribution. Such a fit is a solution to the constrained optimisation problem: 
\begin{equation}
    (\hat p,\hat\eta)=\operatorname*{argmin}_{p,\eta}\,\ell(p,\eta).
\end{equation}
where $\ell$ is a negative-log-likelihood function on the predicted values for the observables incorporating the constraint that the nuisance parameters $\eta$ cannot vary significantly outside of their confidence regions. In \hi~, we model our theoretical predictions by a function $f(p,\eta)$, and we define the log-likelihood function:
\begin{equation}
    \ell(p,\eta)=\ell_{\mathcal O}(p,\eta)+\ell_{\eta}(\eta),
\end{equation}
where the first term corresponds to the log-likelihood of our prediction for the observables and the second term is a penalisation term that blows up whenever the nuisance parameters take values outside of their confidence intervals. The prediction likelihood is modelled as a joint copula distribution of the residues $r(p,\eta)=f(p,\eta)-\mathcal{O}_{\rm exp}$, and the nuisance penalisation likelihood is modelled as a separate copula distribution of the nuisances themselves, hence peaked at the central values of each. \smallbreak
Then we implemented two distinct algorithms to manage the nuisance term and reduce the likelihood to a function $\tilde\ell(p)$ of the fit parameters only.
The first one consists in profiling the general likelihood over the nuisances to find the best available combination of values for the nuisance parameters for a given set of fit parameters $p$:
\begin{equation}
    \tilde\ell(p)=\min_{\eta}\ell(p,\eta).
\end{equation}
The second one consists in marginalising the nuisance parameters by computing the (Gaussian) covariance matrix $\Sigma_{\rm th}$ of the model prediction vector in the SM\footnote{Deviation of the model covariance from the SM has been tested to be small.}, adding it to the covariance matrix of the experimental data that is used to define the copula entering the expression of $\ell_{\mathcal O}$:
\begin{equation}
    \tilde\ell(p)=\eval{\ell_{\mathcal{O}}(p,\eta_0)}_{\Sigma=\Sigma_{\rm exp}+\Sigma_{\rm th}}.
\end{equation}
Compared with the legacy \spi~approximation, where theory uncertainties were estimated from a first-order Taylor expansion of each observable, this mode keeps the full configured nuisance distributions at the sampling stage and only makes a Gaussian approximation at the level of the induced observable covariance.  It is therefore well suited for global scans and graphical-interface best-fit studies: the expensive nuisance propagation is amortised in the covariance construction, and the subsequent minimisation involves only the parameters of interest.  It is complementary to the profiled-nuisance likelihood described above, which retains a more explicitly frequentist interpretation but is more expensive when the number of nuisance parameters is large.\smallbreak
We use \texttt{MINUIT2}~\cite{James:1975dr} to perform the minimisation of the log-likelihood and find the best fit point $\hat p$ along with the full covariance matrix around $\hat p$. Then, one can compute a 2-dimensional confidence contour for a given $z$ value. For this, we make use of the Wilks approximation, stating that near the best fit point the likelihood behaves like a $\chi^2$ distribution, and we define the confidence contour as the set of points in 2-d space such that:
\begin{equation}
    \Delta\ell(p_1,p_2)=\bar\ell(p_1,p_2)-\ell(\hat p)=\Delta\chi^2_n(z),
\end{equation}
where $\bar\ell(p_1,p_2)$ is a new log-likelihood function in which all fit parameters $p_\perp$ orthogonal to the $(p_1,p_2)$ plane have been integrated in some way, and $\Delta\chi_n^2(z)$ is the $n$-degrees of freedom chi-squared variation corresponding to a $z\sigma$ deviation from the best fit value. In \hi~we implemented three ways of integrating the orthogonal parameters: 
\begin{enumerate}
    \item \texttt{SLICE}: The orthogonal parameters are fixed to their best-fit value, i.e. $\bar\ell(p_1,p_2)=\ell(p_1,p_2,\hat p_\perp)$.
    \item \texttt{FREE\_PROJECTION}: The likelihood is profiled over the orthogonal parameters which are let completely free, i.e. $\bar\ell(p_1,p_2)=\min_{p_\perp}\ell(p_1,p_2,p_\perp)$.
    \item \texttt{PRIOR\_CONSTRAINED\_PROJECTION}: The likelihood is profiled over the orthogonal parameters which are constrained by a gaussian distribution built using the best fit information, i.e. $\bar\ell(p_1,p_2)=\min_{p_\perp}\qty[\ell(p_1,p_2,p_\perp)+\ell_\perp(p_\perp)]$ and $\ell_\perp(p_\perp)=(p_\perp-\hat p_\perp)^T\Sigma_\perp(p_\perp-\hat p_\perp)$.
\end{enumerate}
Another usual way of dealing with the orthogonal fit parameters is to fix them to their SM value, but this amounts to performing a 2-d fit only on $(p_1,p_2)$ in the first place, therefore it is not implemented separately in \hi. \smallbreak
There are two contouring algorithms implemented in \hi: the first one uses the already existing algorithm in \texttt{MINUIT2} and the second one is an adaptive marching squares (AMS) algorithm. Although \texttt{MINUIT2}'s algorithm is usually faster, it is targeted at ellipsoidal contours around best-fit points which are not too far from the quadratic approximation. On the other hand, the general-purpose AMS algorithm is usually slower but works well for any contour topology.\smallbreak
We demonstrate in Fig. \ref{fig:contours_proj} the result of these three projections. We performed a 4-dimensional fit of new physics contributions $\delta C_9,\delta C_{10},\delta C_{Q_1},\delta C_{Q_2}$ to the CMS measurements of the angular distributions of  the $B^0\to {K^*}^0\mu^+\mu^-$ decay \cite{CMS:2024atz}. The best fit point in 4-d space is found to be coherent with \cite{Hurth:2025vfx}. As expected, the slice method yields the smallest confidence contour, while the unconstrained profiling yields the largest. The constrained profiling is theoretically bound to sit between the other two, and will be closer to the slice contour the smaller the fit uncertainty is. Interestingly, the enlargement on this particular fit is only found in the right-handed $C_9+C_{10}$ direction, while no enlargement is found in the orthogonal left-handed $C_9-C_{10}$ direction, reflecting the relative steepness of the likelihood in the two directions.

In the public API, this workflow is exposed by registering the selected observables in an \texttt{ObservableInterface} and passing it to the statistical layer together with the nuisance configuration.  The same observable vector can then be used for a central-value prediction, a Monte-Carlo theory covariance estimate, or a full profiled-likelihood calculation.  This avoids duplicating observable definitions between prediction and fitting workflows.

\subsection{Calculation workflow}

The calculation workflow in \hi~is modular.  Short-distance effects are encoded in Wilson coefficients, as introduced in Section~\ref{sec:user_guide} and in the effective-Hamiltonian examples above.  The \texttt{PhysicalModel} layer manages the model-dependent short-distance part, while the \texttt{BusinessLogic} layer implements decay-specific long-distance quantities, observable projections and numerical integrations.  A typical computation therefore follows three steps: first the Wilson coefficients are computed or imported at the matching scale; second, the requested coefficients are evolved to the scale used by the decay calculator; third, observables and statistical quantities are evaluated from the shared runtime parameter database.

This framework accommodates both established scenarios, such as the SM, THDM and SUSY models implemented natively, and new BSM models interfaced through \mty~at leading order.  In all cases, the observable and statistical layers receive the same coefficient objects, which keeps validation and interface behaviour independent of the origin of the short-distance input.

\section{Validation and benchmarks}\label{sec:validation}

The validation of \hi~has two main goals: to compare the native hard-coded calculations with \spi~for the common SM, THDM and SUSY benchmarks, and to compare the native route with the \mty-based route on models available in both workflows.

\subsection{Validation of \spi~legacy}

In order to validate the new \hi~framework, we first compared the hard-coded calculations with \spi~for the models available in the legacy code.  Since the model dependence enters primarily through the short-distance coefficients, the first validation step is a direct comparison of Wilson coefficients at the matching scale.  The long-distance observables are then checked separately in the SM, where the calculation is faster and the observable implementation is model independent once the Wilson coefficients are fixed.

Table~\ref{tab:comparison_superiso_B} shows the comparison for the standard \(B\)-sector coefficients.  The SM, THDM and SUSY benchmarks are displayed as stacked blocks rather than parallel column groups, which keeps the numerical entries readable while preserving the same information.  The full set of coefficients used in the validation, including scalar and primed operators, is reported in \ref{app:extended_validation_tables}.  The relative differences remain below the per-mille level in the displayed benchmark, confirming that the native \hi~implementation reproduces the corresponding \spi~short-distance calculation.  Additional comparisons were performed after QCD running to the low-energy scale and showed agreement with \spi~within numerical precision.

\begin{table}[!ht]
\centering
\setlength{\tabcolsep}{4pt}
\renewcommand{\arraystretch}{1.1}
\begin{adjustbox}{max width=\textwidth}
\begin{tabular}{lSSS}
\hline
\textbf{coefs} & \textbf{SI} & \textbf{HI} & \textbf{\%} \\
\hline
\multicolumn{4}{c}{\(\mu_W=\mathcal{O}(M_W)\)} \\
\hline
\multicolumn{4}{c}{\textbf{SM}} \\
\hline
C1 & 0.1583963453 & 0.1583963451 & 0 \\
C2 & 1.001843983 & 1.001843983 & 0 \\
C3 & -0.0002931687782 & -0.0002931687782 & 0 \\
C4 & -0.003248811761 & -0.003247432028 & 0.042478 \\
C5 & 3.386474459e-05 & 3.386879571e-05 & 0.011962 \\
C6 & 6.349639566e-05 & 6.349639566e-05 & 0 \\
C7 & -0.2107427536 & -0.2106774232 & 0.031005 \\
C8 & -0.1151167202 & -0.1150915551 & 0.021863 \\
C9 & 1.930302269 & 1.92962599 & 0.035041 \\
C10 & -4.171695742 & -4.168626236 & 0.073606 \\
\hline
\multicolumn{4}{c}{\textbf{THDM}} \\
\hline
C1 & 0 & 0 & 0 \\
C2 & 0 & 0 & 0 \\
C3 & -3.723792155e-08 & -3.723792155e-08 & 0 \\
C4 & 9.296889404e-06 & 9.296896376e-06 & 0.000075 \\
C5 & 1.400574929e-08 & 1.400577667e-08 & 0.000195 \\
C6 & 2.626083695e-08 & 2.626083695e-08 & 0 \\
C7 & 0.0008007874546 & 0.0008004786918 & 0.038565 \\
C8 & 0.0008704866784 & 0.0008704864666 & 0.000024 \\
C9 & -0.0002778553141 & -0.0002778544545 & 0.000309 \\
C10 & -0.004887954445 & -0.004887955214 & 0.000016 \\
\hline
\multicolumn{4}{c}{\textbf{SUSY}} \\
\hline
C1 & -5.594240306e-07 & -5.594240306e-07 & 0 \\
C2 & 0 & 0 & 0 \\
C3 & 4.9344336e-07 & 4.934424585e-07 & 0.000183 \\
C4 & 4.003675142e-06 & 4.003666186e-06 & 0.000224 \\
C5 & -4.44468528e-08 & -4.444694295e-08 & 0.000203 \\
C6 & -8.333807436e-08 & -8.333798422e-08 & 0.000108 \\
C7 & 0.0008455493107 & 0.0008451479372 & 0.047480 \\
C8 & -0.0101108152 & -0.010110883 & 0.000671 \\
C9 & 0.00588099578 & 0.005879457677 & 0.026157 \\
C10 & 0.005718707991 & 0.00571869065 & 0.000303 \\
\hline
\end{tabular}
\end{adjustbox}
\caption{Comparison between \spi~(SI) and \hi~(HI) for the standard \(B\)-sector Wilson coefficients at matching scale.  The three benchmark models are displayed in stacked blocks to improve readability.  The complete table, including scalar and primed coefficients, is given in \ref{app:extended_validation_tables}.}
\label{tab:comparison_superiso_B}
\end{table}

We compared the Wilson coefficients in all the models common to \spi~and \hi, namely the SM, THDM and SUSY benchmarks, using identical input parameters.  This establishes the consistency of the native \hi~implementation for the hard-coded legacy scenarios.

\begin{figure}[t]
    \centering
    \includegraphics[width=.499\linewidth]{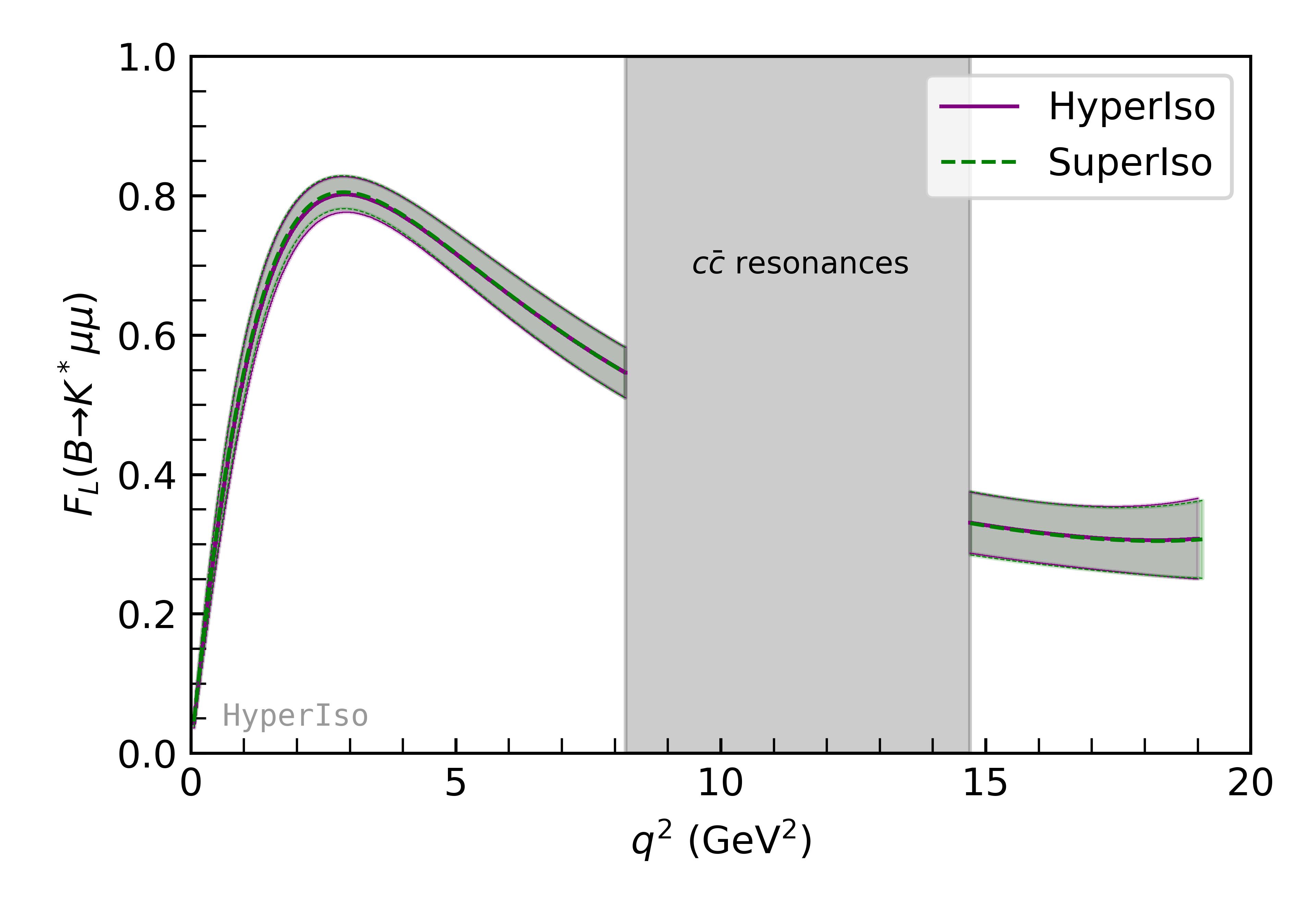}\hfill
    \includegraphics[width=.499\linewidth]{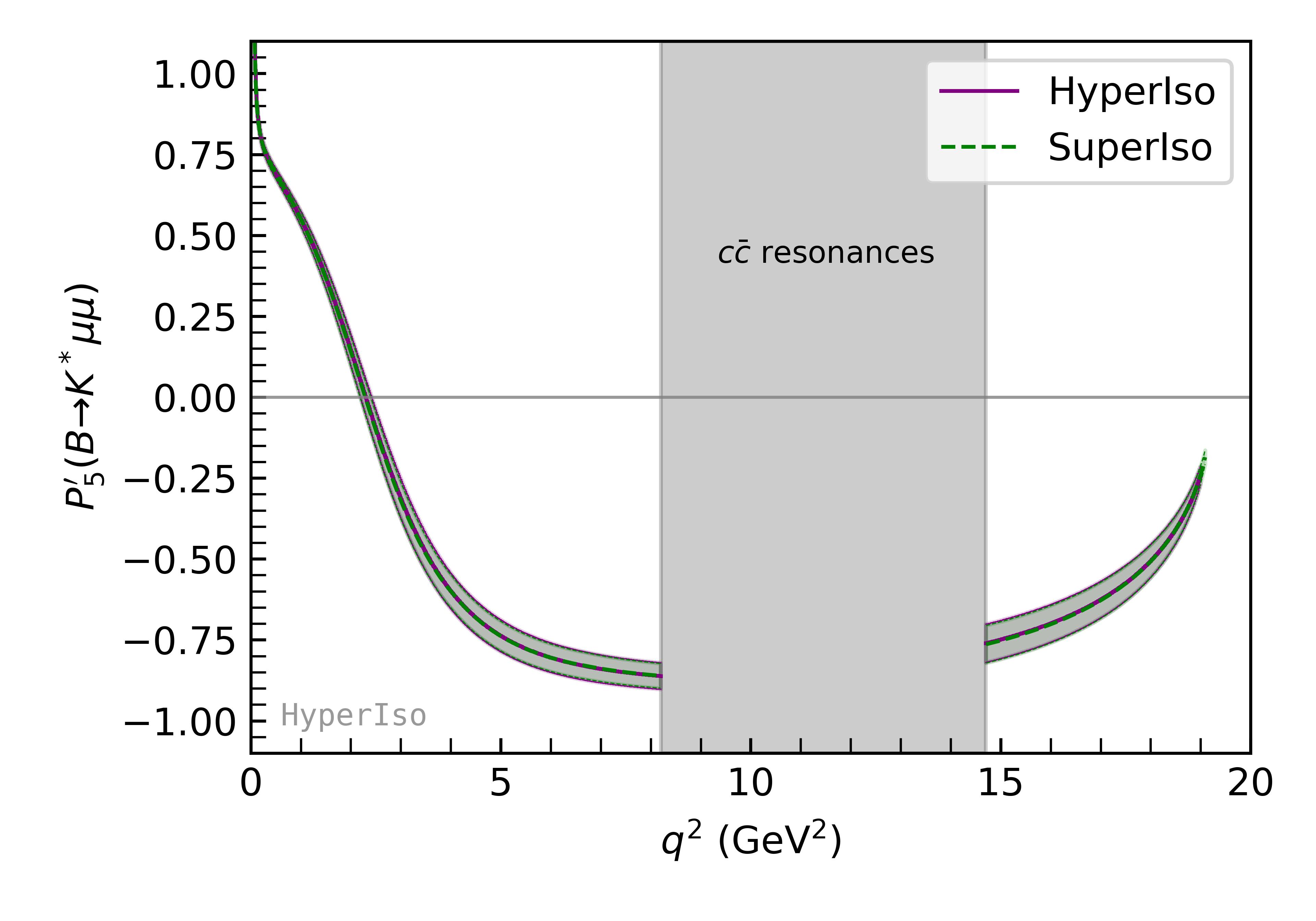}
    \caption{Comparison between \spi~and \hi~for some $B\to K^*\mu\mu$ angular observables.}
    \label{fig:hyp_sup_comparison_BKstarGamma}
\end{figure}

We next compared the long-distance implementation. This being model independent, we only made the comparison in the SM. An example is given in Fig.~\ref{fig:hyp_sup_comparison_BKstarGamma} for the angular observables $F_L$ and $P'_5$ of the $B\to K^*\mu\mu$ semileptonic decay as a function of the transferred momentum $q^2$. Excellent agreement is found for the central value and the uncertainty bands. This validates both the implementation of the observables in \hi~and the linear approximation made in \spi~for the estimation of uncertainties.

\begin{figure}[ht]
    \centering
    \includegraphics[width=0.7\linewidth]{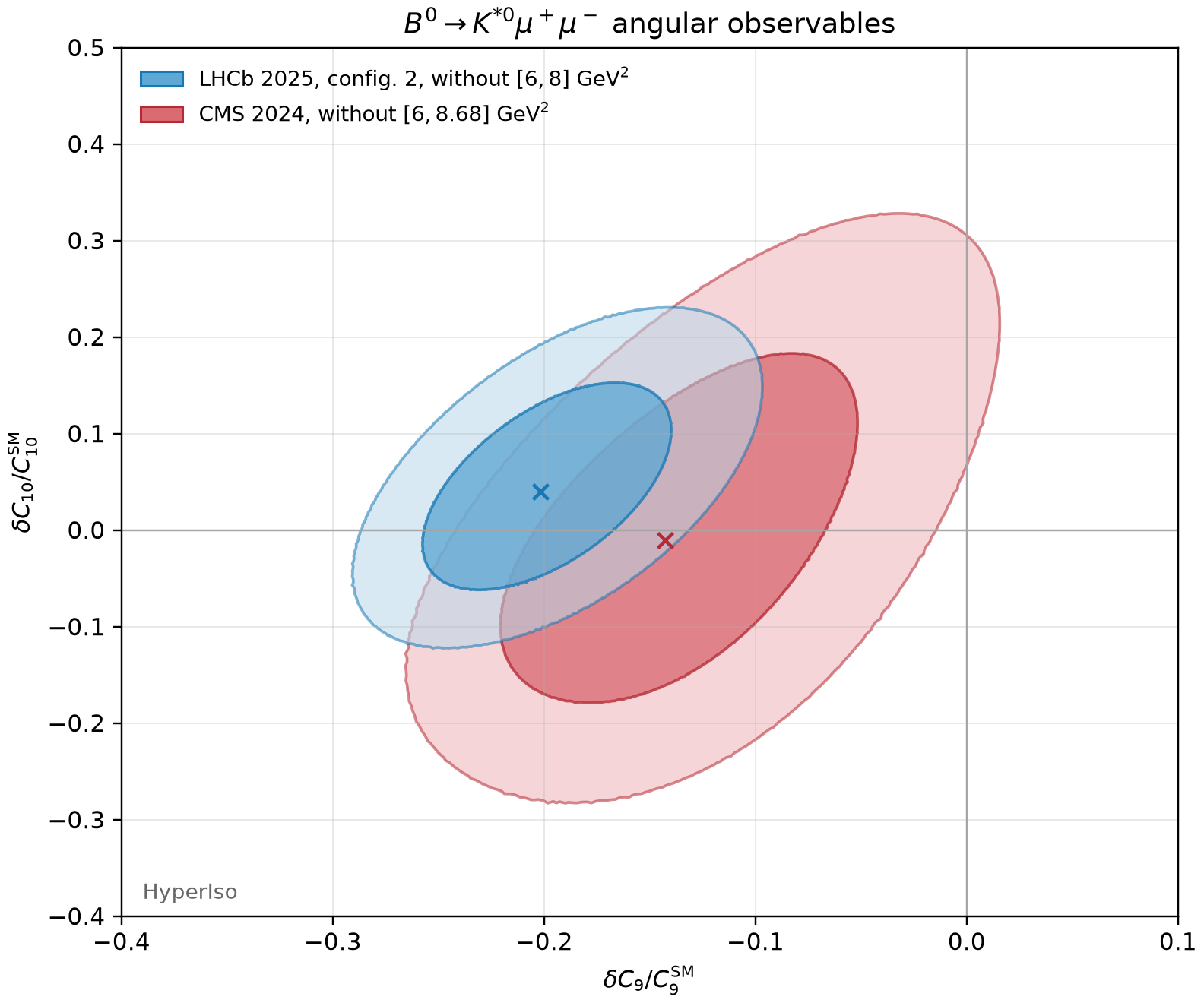}
    \caption{Model-independent ${C_9, C_{10}}$ fit to angular $B^0\to {K^*}^0\mu^+\mu^-$ observables, using the measurements from CMS \cite{CMS:2024atz} and LHCb \cite{LHCb:2025mqb}.}
    \label{fig:2dfitsang}
\end{figure}

Finally, we tested the fitting and contouring features of \hi{}. We reproduced the model-independent new physics fits of $\delta C_9,\delta C_{10}$ to angular observables of the $B^0\to {K^*}^0\mu^+\mu^-$ decay of \cite{Hurth:2025vfx} using the experimental data from both CMS \cite{CMS:2024atz} and the 2025 update of LHCb \cite{LHCb:2025mqb}. We used the theoretical covariance based likelihood function. The results are shown in Fig.~\ref{fig:2dfitsang} and are in remarkable agreement with \spi{} both regarding the best-fit point and the confidence contours. It should also be noted that producing such results is very quick in \hi{} compared to \spi{}. The plot in Fig.~\ref{fig:2dfitsang} was produced in approximately 5 minutes on 30 threads, using \num{1500} Monte-Carlo draws for the calculation of the theory covariance and a contour resolution of 100.

\subsection{Validation of the \mty~interface}

The second validation step is internal to \hi~and tests the agreement between the native hard-coded Wilson-coefficient implementation and the generic \mty-based route.  We perform this comparison on the SM and on a type-II THDM benchmark, using the same LHA input and the same normalisation conventions.  The comparison is made at the matching scale, where the Wilson coefficients are produced either by the native analytical implementation or by the \mty~workflow.  This is the relevant scale for validating the interface itself: after matching, the QCD evolution to the hadronic scale is performed by the same \hi~running code in both workflows.  The running part has already been validated against \spi~in the previous subsection.

The present \mty~interface is used for the leading-order calculation.  We therefore restrict the direct native--\mty~comparison to LO coefficients.  Higher-order QCD corrections are not generated by \mty~in this workflow; they are instead part of the native \hi~implementation and of the common QCD-running stage.  In Table~\ref{tab:comparison_marty_B} (and in the extended Tables~\ref{tab:comparison_marty_full_sm} and~\ref{tab:comparison_marty_full_bsm}), we show only coefficients that are non-zero in at least one of the compared SM or type-II THDM BSM entries, after removing entries that are redundant for the displayed benchmark.  In particular, the printed tables keep the muonic scalar coefficients but omit the electron and tau scalar entries, and they do not list the full set of \(C^\prime_Q\) coefficients.  This keeps the comparison focused on the operators used in the reference phenomenological channels while the complete machine-readable export is kept in the validation files.

For the MARTY route, the displayed BSM entries are obtained from the
diagrammatically selected non-SM contribution. They are not constructed by
subtracting a separately evaluated SM coefficient from a full-model result.
The total coefficient is obtained by adding the selected SM and genuine BSM
components.
\begin{table}[!t]
\centering
\small
\setlength{\tabcolsep}{3pt}
\renewcommand{\arraystretch}{1.08}
\begin{adjustbox}{max width=1.0\textwidth}
\begin{tabular}{lcccccc}
\hline
\textbf{Coefficient} & \multicolumn{3}{c}{\textbf{SM}} & \multicolumn{3}{c}{\textbf{THDM type-II BSM}} \\
 & \textbf{HI} & \textbf{\mty} & \(\boldsymbol{\delta_{\rm rel}}\) & \textbf{HI} & \textbf{\mty} & \(\boldsymbol{\delta_{\rm rel}}\) \\
\hline
\multicolumn{7}{c}{\(\mu_W=\mathcal{O}(M_W)\), matching scale, LO} \\
\hline
\(C_{2}\) & \num{1} & \num{1.000064} & \num{6.4e-05} & \num{0} & \num{0} & \multicolumn{1}{c}{--} \\
\(C_{7}\) & \num{-0.19411475} & \num{-0.1942087} & \num{0.000484} & \num{-0.099068547} & \num{-0.0990574} & \num{0.000113} \\
\(C_{8}\) & \num{-0.096902814} & \num{-0.09697558} & \num{0.000751} & \num{-0.099275045} & \num{-0.09925882} & \num{0.000163} \\
\(C_{9}\) & \num{2.0076248} & \num{2.0076248} & \num{0} & \num{0.00012406279} & \num{0.0001228477} & \num{0.00979} \\
\(C_{10}\) & \num{-4.4637361} & \(\num{-4.479714}+\num{0.08860065}\,i\) & \num{0.0202} & \num{-0.0065801945} & \num{-0.006560991} & \num{0.00292} \\
\hline
\end{tabular}
\end{adjustbox}
\caption{Native-\hi/\mty~comparison for the non-zero standard \(B\)-sector Wilson coefficients at matching scale.  Only the SM and genuine type-II THDM contributions are displayed, since the total THDM coefficient is obtained by their sum.  The last column in each block gives the relative difference defined in Eq.~\eqref{eq:marty_relative_difference}.  The complete displayed table is given in \ref{app:extended_validation_tables}.}
\label{tab:comparison_marty_B}
\end{table}
It is important to note that the \mty-based route used here should not be interpreted as a one-to-one implementation of the dedicated matching procedure employed in \hi~or in \spi.  In the current interface, \mty~generates amplitudes which are then projected directly onto the effective-operator basis used by \hi.  This direct projection is sufficient to validate the normalisation, basis conventions and the generic-model data flow, but it is not yet equivalent to a dedicated EFT matching calculation with all the convention-dependent rearrangements, projections and operator reductions implemented at the same level as in the native code.  Consequently, small but visible differences can appear for some coefficients, especially in sectors where the projection is numerically sensitive or where scalar and primed structures are involved.  A dedicated matching layer in \mty~is under development; once available, it is expected to provide a closer analogue of the native \hi/\spi~matching workflow and may be included in a future release.

Table~\ref{tab:comparison_marty_B} summarises the numerical comparison for the non-zero standard \(B\)-sector coefficients.  The complete set of displayed matching-scale entries, including the selected primed and scalar operators, is reported in \ref{app:extended_validation_tables}.  For each coefficient, we display only the SM contribution and the genuine type-II THDM contribution.  The total THDM coefficient is not shown as an independent column because it is obtained by summing the two displayed contributions.  We report the relative difference

\begin{equation}
  \delta_{\rm rel}(C_i)=\frac{\left|C_i^{\rm HI}-C_i^{\rm MARTY}\right|}{\left|C_i^{\rm HI}\right|},
  \label{eq:marty_relative_difference}
\end{equation}
whenever the native coefficient is non-zero; a dash denotes entries for which this normalisation is not defined. An interesting feature is the appearance of nonzero, yet small imaginary parts in MARTY's calculation. The reason for this is twofold: first, exact CKM unitarity is not enforced numerically as we rely on the Wolfenstein parameterisation, and second, the light quark masses ($c$ and $u$) are not set to zero but rather kept at their physical value in the $\overline{\rm MS}$ renormalisation scheme, leading to very good yet not exact GIM cancellation, and the appearance of nonzero imaginary parts as a distinctive feature of this approximation being relaxed.  

For MARTY-backed evaluations, each numerical evaluation also receives a
private runtime directory containing its parameter input and Wilson-coefficient
output files. Generated sources, libraries and numerical executables remain
shared read-only cache artefacts. Their generation or invalidation is
serialised, while already compiled numerical evaluations can run concurrently.
The per-evaluation runtime directory is removed after its output has been
consumed.

\section{Performance and scaling}\label{sec:performance}

Unless otherwise stated, the benchmarks use \hi~\releaseversion{}
(commit \releasecommit), a GCC 14.3.0 Release build
(\texttt{-O3 -DNDEBUG}) configured with CMake 3.28.3, on an Intel Core
i9-14900K under Ubuntu 24.04.4 LTS on WSL2. The environment exposed 32 logical
CPUs and 31~GiB of RAM (\texttt{x86\_64}; Linux kernel
\path{6.18.33.1-microsoft-standard-WSL2}). The requested thread count and the
number of timed repetitions are reported for each benchmark.

The computational cost of \hi~is not uniform across the available observables.  Wilson coefficients are evaluated once for a chosen model, perturbative order and scale configuration, whereas the long-distance part is organised by decay channel and may involve bin-by-bin integrations, form-factor evaluations, QCD factorisation terms and angular projections.  For a selected set $S$ of observables grouped into decay families $d$, the wall time of a call to \texttt{ObservableInterface::compute\_all()} may be schematically written as:
\begin{equation}
    T_{\rm obs}(S)\simeq T_{\rm WC}
    +\sum_{d}\left[T_{\rm cache}^{(d)}
    +N_{\rm bin}^{(d)}N_q^{(d)}C_{\rm amp}^{(d)}
    +N_{\rm obs}^{(d)}C_{\rm proj}^{(d)}\right],
    \label{eq:obs_complexity}
\end{equation}
where $T_{\rm WC}$ is the Wilson-coefficient setup cost, $T_{\rm cache}^{(d)}$ denotes decay-specific cached quantities, $N_{\rm bin}^{(d)}$ is the number of selected $q^2$ bins, $N_q^{(d)}$ is the effective number of quadrature points per bin, $C_{\rm amp}^{(d)}$ is the cost of evaluating the amplitudes or form-factor combinations entering the integrand, and $C_{\rm proj}^{(d)}$ is the comparatively small cost of projecting the cached angular coefficients onto the requested observables.  This separation is useful in practice because most of the speed-critical work is decay-local: once a decay cache has been filled, many observables can be extracted from the same intermediate quantities.

\begin{figure}[!ht]
    \centering
    \includegraphics[width=0.8\textwidth]{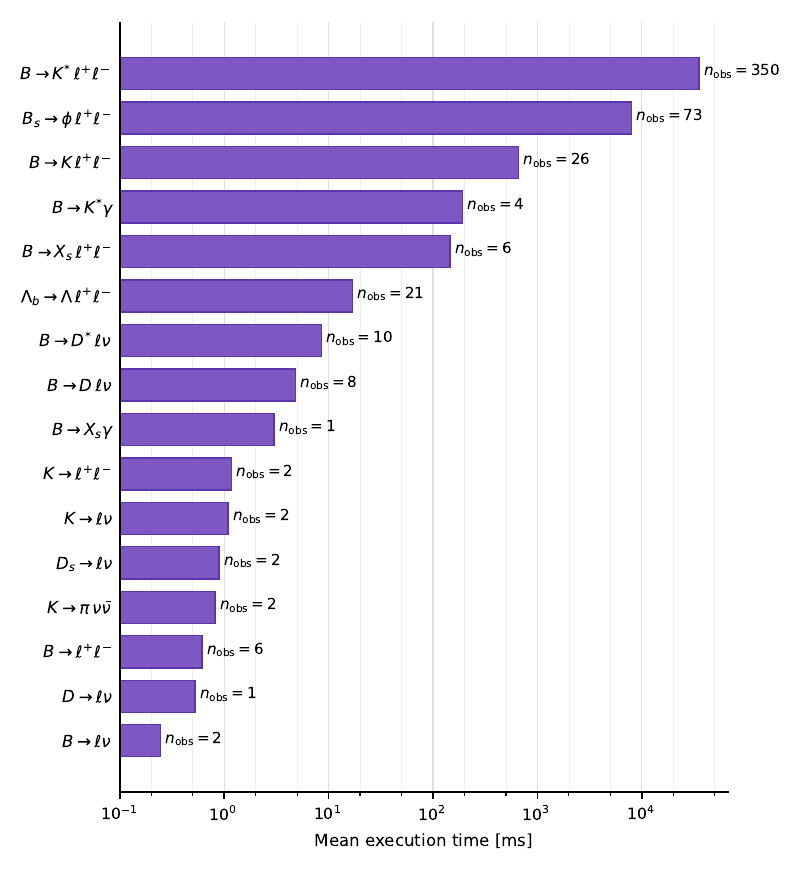}
    \caption{Per-decay benchmark of the mean wall time of \texttt{ObservableInterface::compute\_all()} at NLO.  The benchmark uses 30 timed repetitions after warm-up calls and displays the number of observable entries associated with each decay family.  Since all channels are measured on the same setup, the figure is intended primarily as a relative cost comparison; absolute timings depend on the compiler, CPU and runtime configuration.}
    \label{fig:decay_benchmark}
\end{figure}

This hierarchy is visible in the per-decay benchmark shown in Fig.~\ref{fig:decay_benchmark}.  Each entry reports the mean time of one \texttt{compute\_all()} call at NLO over 30 timed repetitions after warm-up calls, using the same input point. The horizontal axis is logarithmic because the dynamic range is very large: the fastest leptonic modes are below the millisecond scale, whereas the most expensive binned semileptonic modes can require seconds to tens of seconds.  The dominant entries are the angular decays with many binned observables, in particular $B\to K^*\ell\ell$ and $B_s\to\phi\ell\ell$.  The former contains 350 observable entries in this benchmark and is more than four orders of magnitude slower than the simplest leptonic channels, even after using the threaded implementation.  This behaviour is expected from Eq.~\eqref{eq:obs_complexity}: the cost is driven not only by the number of published observables but also by the shared amplitude and QCD-factorisation evaluations needed for each bin.

The largest gains are therefore obtained by optimising the heavy decay caches rather than the lightweight observable projections.  In the $B\to K^*\ell\ell$ implementation, the binned angular observables are expressed through a common set of angular coefficients $J_i$.  These coefficients are integrated in a single pass per bin and per CP conjugation using a fixed Gauss--Legendre rule.  At each quadrature point the transversity amplitudes are evaluated once, and the set of $J_i$ combinations is accumulated simultaneously.  This avoids the repeated-integral pattern in which many observables would independently re-evaluate the same amplitudes.  In addition, the expensive QCD-factorisation lookup tables entering the amplitudes are filled in parallel.  The lookup grid is split into contiguous chunks over worker threads; each worker owns local form-factor and QCD-factorisation calculator instances and writes to disjoint cache entries.  Apart from exception collection, the inner loop is therefore lock-free.  For a lookup size $N_{\rm lookup}$ and $n_t$ worker threads, the cache-filling contribution scales approximately as:
\begin{equation}
    T_{\rm cache}^{B\to K^*\ell\ell}(n_t)
    \simeq
    \frac{N_{\rm lookup} C_{\rm QCDf}}{\min(n_t,N_{\rm lookup})}
    + T_{\rm launch}+T_{\rm memory},
    \label{eq:bkstar_cache_scaling}
\end{equation}
up to load imbalance, thread-launch overhead and memory-bandwidth effects. The number of threads allocated to this parallelisation step can be set in the individual \texttt{DecayConfig} structures of the relevant decays.

\begin{figure}[!t]
    \centering
    \includegraphics[width=0.95\textwidth]{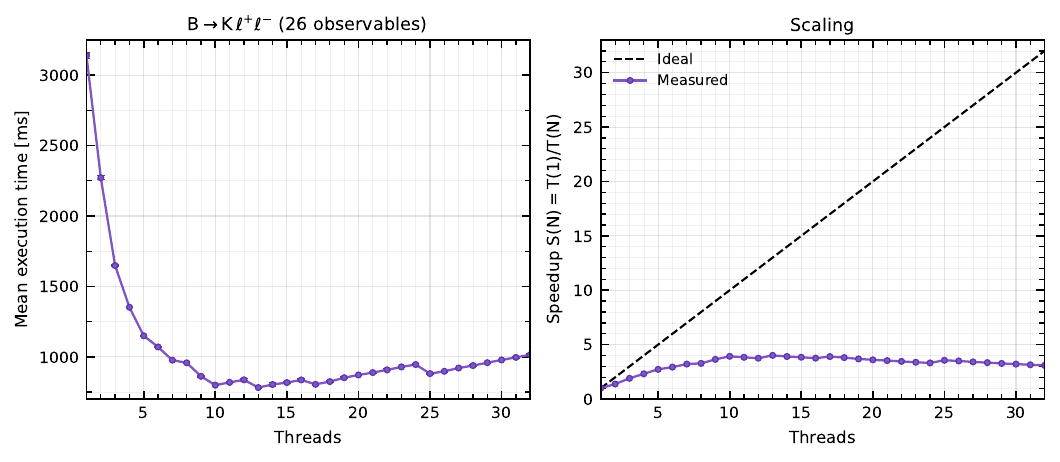}
    \caption{Thread-scaling benchmark for a binned semileptonic \texttt{compute\_all()} workload at NLO.  The left panel shows the mean wall time per call as a function of the requested number of threads.  The right panel shows the measured speedup relative to one thread and the ideal linear behaviour.  The saturation beyond approximately 15--20 threads indicates the transition from amplitude/QCD-factorisation dominated work to serial overheads and memory effects.}
    \label{fig:bkstarll_thread_scaling}
\end{figure}

Figure~\ref{fig:bkstarll_thread_scaling} shows the corresponding thread-scaling study for a representative binned semileptonic benchmark.  The one-thread runtime is about $3.79\,\mathrm{s}$ per \texttt{compute\_all()} call.  Increasing the number of worker threads gives a rapid reduction at small thread counts, reaching a speedup of about $4.2$ at 8 threads and about $5.4$ at 20 threads for $B\to K^*\ell^+\ell^-$, and maximum 4 at 10 threads for $B\to K\ell^+\ell^-$.  The best point in the supplied scan is around 25 threads, with a mean time of about $0.64\,\mathrm{s}$, corresponding to a speedup close to 6.  Beyond roughly 15--20 threads the curve flattens: the remaining serial work, cache construction outside the parallel region, memory traffic and thread scheduling overhead dominate over the parallel QCD-factorisation work.  The measured curve is therefore far from ideal linear scaling at large $n_t$, but the reduction by a factor of five to six is sufficient to make binned scans and repeated fits substantially more practical.

The same considerations determine the cost of statistical analyses.  In the $\chi^2$ backend based on a Monte-Carlo theory covariance, the nuisance parameters are sampled once to construct the Monte-Carlo theory covariance, after which the fit itself is performed in the lower-dimensional space of the parameters of interest.  If $D$ observables enter the fit, $N_{\rm MC}$ accepted nuisance samples are used for the covariance estimate, and $N_{\rm eval}$ model evaluations are required by the minimiser, the leading cost is:

\begin{equation}
    T_{\chi^2}\simeq
    N_{\rm MC}\,T_{\rm pred}(D)
    +\mathcal{O}(N_{\rm MC}D^2)
    +\mathcal{O}(D^3)
    +N_{\rm eval}\,T_{\rm pred}(D),
    \label{eq:chi2_complexity}
\end{equation}
where $T_{\rm pred}(D)$ is the time needed to evaluate the selected observable vector.  The second term is the empirical covariance construction, the third is the inversion of the total covariance matrix, and the last term is the deterministic minimisation.  The memory usage is $\mathcal{O}(N_{\rm MC}D+D^2)$ when all accepted samples are kept for diagnostics or written to CSV.  Consequently, for global fits involving expensive binned observables, the dominant scaling variable is often the cost of repeated observable prediction rather than the linear algebra.

For this reason, \hi~uses two complementary levels of parallelism.  The decay-level parallelism described above remains the most efficient acceleration mechanism for a single expensive prediction, and it is used by deterministic stages such as central-value evaluations, likelihood minimisation, Hessian or profile scans, and contour construction.  However, Fig.~\ref{fig:bkstarll_thread_scaling} also shows that this inner parallelism is not asymptotically linear: once the QCD-factorisation part has been distributed over enough workers, the residual serial work, memory traffic and scheduling overhead set a practical speedup plateau.  The Monte-Carlo covariance stage has a different structure.  The accepted nuisance draws are statistically independent which makes them good candidates for parallelisation.  In the multi-threaded Monte-Carlo mode, \hi~therefore gives priority to this outer parallelism: each worker owns an independent observable model and parameter runtime context, draws nuisance points from the shared sampler in a synchronised way, and evaluates its assigned predictions without modifying the global parameter state.  To avoid nested over-subscription, the internal decay thread count is temporarily reduced, by default to one, while the Monte-Carlo workers are running.  After the covariance sample has been filled, the original decay-thread configuration is restored automatically and the subsequent deterministic stages again use the decay-local threaded caches. The number of threads allocated to Monte-Carlo sampling is set by the \texttt{MC\_threads} parameter of the \texttt{StatisticConfig}. 

With this scheduling policy, the covariance-building part of Eq.~\eqref{eq:chi2_complexity} is better represented as:
\begin{equation}
    T_{\rm cov}(n_{\rm MC})
    \simeq
    \frac{N_{\rm MC}}{n_{\rm MC}}\,
    T_{\rm pred}(D; n_t=1)
    +\mathcal{O}(N_{\rm MC}D^2)
    +T_{\rm sched},
    \label{eq:mc_parallel_covariance}
\end{equation}
where $n_{\rm MC}$ is the number of Monte-Carlo workers and $T_{\rm sched}$ collects sampler synchronisation, worker setup and result-merging overheads.  Since the dominant term is usually the prediction time, the scaling of the covariance stage is close to linear in $n_{\rm MC}$ until the number of active workers approaches the available hardware parallelism or the bookkeeping terms become comparable to the prediction cost.  The full workflow therefore uses the parallelism at the level where it is most effective: sample-level parallelism for the Monte-Carlo covariance construction, and decay-level parallelism for the remaining repeated predictions required by fits and scans.

\section{Software availability and reproducibility}\label{sec:software_availability}

The version of \hi\ described in this article is the public release
\texttt{\releaseversion}, released on \releasedate. The source code,
Python distributions, documentation and frozen reproducibility references
are available from:

\begin{itemize}

    \item \hi{} website:
          \url{https://hyperiso.in2p3.fr/};
    \item source release: \texttt{HyperIso \releaseversion};
    \item release tag commit: \releasecommit;
    \item frozen-reference provenance:
          \path{reproducibility/expected_outputs/reference_metadata.json},
          which records the reference-build source commit and executable SHA-256;
    \item developer repository:
          \url{https://github.com/Hyperiso/Hyperiso};
    \item GitHub release:
          \url{https://github.com/Hyperiso/Hyperiso/releases/tag/\releaseversion};
    \item version-specific software archive:
          \softwaredoi;
    \item Python package:
          \url{https://pypi.org/project/pyhyperiso/\packageversion/};
    \item API documentation:
          \url{https://hyperiso.github.io/Hyperiso/};
    \item license: GNU General Public License, version 3 or later.
\end{itemize}

The recommended installation route for the Python interface is:

\begin{terminal}
python -m pip install "pyhyperiso==1.0.3"
\end{terminal}

The release contains Linux x86-64 manylinux wheels for CPython 3.10--3.12
and a source distribution. Other platforms can build the C++, CLI and Python
interfaces from source.

The standard CMake workflow for the C++ and CLI interfaces is:
\begin{terminal}
cmake -S Hyperiso/Hyperiso/core -B build \
  -G Ninja \
  -DCMAKE_BUILD_TYPE=Release \
  -DBUILD_WITH_CLI=ON \
  -DBUILD_WITH_PYTHON=ON
cmake --build build --parallel
cmake --install build --prefix "$HOME/.local"
\end{terminal}
Here \path{-DBUILD_WITH_CLI=ON} adds the standalone \texttt{hyperiso-ui}
executable to the standard library build. The separate option
\path{-DBUILD_WITH_APP=ON} enables the development applications and benchmark
executables distributed inside the C++ modules.

The generic \mty-based workflow requires an external \mty~installation.  Native examples require GSL and a C++20-compatible compiler.  Optional spectrum-generation workflows may also rely on \texttt{2HDMC}~\cite{Eriksson:2009ws} or \texttt{SOFTSUSY}~\cite{Allanach:2001kg}.

The release archive contains a root-level \texttt{reproducibility/} directory.  This directory is the canonical location for the frozen input files, run scripts, expected text outputs and numerical tolerances used to reproduce the compact reference examples reported in this paper.  The suite is intentionally command-line based: the CLI provides the most stable release-level interface, while the C++ and Python examples distributed with the source tree expose the same workflows programmatically.

\begin{terminal}
reproducibility/
  README.md
  manifest.json
  inputs/
    sm_reference.flha
    thdm_reference.lha
    susy_reference.slha
  scripts/
    run_cli_suite.sh
    normalize_cli_output.py
    check_expected_outputs.py
    freeze_reference_outputs.py
  expected_outputs/
    reference_metadata.json
\end{terminal}

The suite complements the unit and integration tests; it does not replace them.
It can be executed from the repository root with

\begin{terminal}
HYPERISO_REPRO_THREADS=1 \
  ./reproducibility/scripts/run_cli_suite.sh
\end{terminal}

The reference suite is summarised in Table~\ref{tab:repro_suite}.  Deterministic examples are compared using absolute and relative tolerances.  The Monte-Carlo example is run with a fixed seed through the \texttt{--seed} option and with the serial MC path by default; it is therefore suitable for release-to-release comparisons.  Parallel MC runs are kept for performance studies rather than bitwise reproducibility checks.

\begin{table}[!ht]
\centering
\small
\renewcommand{\arraystretch}{1.15}
\begin{tabularx}{\textwidth}{@{}p{0.08\textwidth}p{0.28\textwidth}X@{}}
\toprule
\textbf{ID} & \textbf{Workflow} & \textbf{Reference check} \\
\midrule
R1 & $B$-sector Wilson coefficients &
\texttt{BCoefficients}; $C_7,C_8,C_9,C_{10}$ at matching and low scales. \\
R2 & Scalar Wilson coefficients &
\texttt{BScalarCoefficients}; $C_{Q_1}^{\mu},C_{Q_2}^{\mu}$ in the scalar sector. \\
R3 & Inclusive and leptonic observables &
\texttt{BR\_Bs\_\_mu\_mu} and \texttt{BR\_B\_\_Xs\_gamma}. \\
R4 & Binned angular observable &
\texttt{F\_L\_B0\_\_K*0\_mu\_mu} in the fixed $q^2$ bin $[1.1,6.0]~\mathrm{GeV}^2$. \\
R5 & Seeded MC uncertainty propagation &
\texttt{statistic summary --uncertainties --draws 200 --seed 123456},
run through the serial MC path, with the accepted-sample CSV checked against
the frozen reference. \\
R6 & THDM archived-spectrum benchmark &
BSM \(B\)-sector coefficients \(C_7,C_8,C_9,C_{10}\) at NNLO, using the
archived THDM spectrum with \texttt{--spectrum true}; no runtime call to
\texttt{2HDMC}. \\

R7 & SUSY archived-spectrum benchmark &
BSM \(B\)-sector coefficients \(C_7,C_8,C_9,C_{10}\) at NNLO, using the
archived SUSY spectrum with \texttt{--spectrum true}; no runtime call to
\texttt{SOFTSUSY}. \\
\bottomrule
\end{tabularx}
\caption{Compact reference reproducibility suite shipped with the release archive.  The commands, input file, expected output names and tolerances are defined in \texttt{reproducibility/manifest.json}.}
\label{tab:repro_suite}
\end{table}

The reference metadata records the source tag and source commit, the executable
SHA-256 digest, input-file hashes, operating system, architecture, compiler,
CMake, GSL, Eigen and Python versions, build type, serial thread count and
measured execution time of each reference case. The timings characterise the
frozen reference host and should not be interpreted as universal performance
values.

\section{Conclusion}\label{sec:conclusion}

We have presented \hi, a new standalone modular program, building on the physics heritage of \spi, designed to evaluate Wilson coefficients, flavour observables and statistical quantities in a unified framework. The C++ core provides the common calculation backend, while the Python, command-line and Dash interfaces make the tool accessible for scripted, automated and interactive workflows. \hi\ combines native implementations for established scenarios with a \mty-based route for user-defined BSM models supported by that interface at leading order, while retaining the internal machinery for observable calculations, uncertainty propagation and statistical interpretation.

The validation against \spi~confirms the consistency of the native calculations for the supported benchmark models. The comparison between the native \hi~ route and the \mty~ route on common benchmark models provides the complementary validation of the generic-model interface; the numerical interpretation of these differences is documented in the validation section and fixed together with the released \mty~benchmark inputs. Together with the user-guide examples and reproducibility package, this makes \hi~ a practical tool for both phenomenological studies and software extensions.

The \releaseversion~release also validates installed-package execution: the statistical
layer resolves nuisance assets through the active runtime provider, and the
public Python API supports exact experimental-measurement selections used by
the released fit-reproduction examples.

\section*{Acknowledgements}
The authors thank S. Neshatpour for insightful discussions. This research is funded in part by the National Research Agency (ANR) under project no. ANR-21-CE31-0002-01. The authors acknowledge the open-source scientific software ecosystem on which \hi~builds, including \mty, \texttt{GSL}, \texttt{pybind11}, \texttt{2HDMC} and \texttt{SOFTSUSY}.  The development of \hi~also benefited from the physics conventions and validation experience accumulated in \spi.

\pagebreak
\appendix

\section{Parameter block structure}\label{app:block_structure}
\subsection{SM Parameters}
\paragraph{Block \texttt{SMINPUTS}}
This block is defined by the SLHA convention. We added two entries to store the redundant parameter $\sin^2\theta_W$ as precise values of this parameter are needed in order to comply with some authors' conventions in observable calculation. Default values are gathered in Table \ref{tab:SMINPUTS}.
\renewcommand{\arraystretch}{1.3}
\begin{table}[!ht]
    \centering
    \begin{tabular}{cccc}
        \hline
        \texttt{ID} & Parameter & Default value \\
        \hline
        \texttt{1}&$\alpha_{\rm em}^{-1}(M_Z)$&\num{127.930}\\
        \texttt{2}&$G_F$&\SI{1.1663785(6)e-5}{\per\gev\squared}\\
        \texttt{3}&$\alpha_s(M_Z)$&\num{0.1180(9)}\\
        \texttt{4}&$M_{Z}^{\rm pole}$&\SI{91.1879(20)}{\gev}\\
        \texttt{5}&$m_b^{\overline{\rm MS}}(m_b)$&\SI{4.186(6)}{\gev}\\
        \texttt{6}&$m_{t}^{\rm pole}$&\SI{172.60(27)}{\gev}\\
        \texttt{7\_1}&$s_w^{2, \overline{\rm MS}}$&\num{0.23122(6)}\\
        \texttt{7\_2}&$s_w^{2, \rm OS}$&\num{0.22348(10)}\\
        \hline
    \end{tabular}
    \caption{Entries of the \texttt{SMINPUTS} block, from PDG \cite{ParticleDataGroup:2026aaa}.}
    \label{tab:SMINPUTS}
\end{table}

\paragraph{Block \texttt{MASS}}
This block follows the SLHA convention and stores particle masses that are not already provided through \texttt{SMINPUTS}. For the light quarks, the default values correspond to running $\overline{\mathrm{MS}}$ masses at the scales indicated below, whereas the charged-lepton and electroweak-boson entries are pole masses. Default values are gathered in Table \ref{tab:MASS}.
\begin{table}[!ht]
    \centering
    \begin{tabular}{ccccc}
        \hline
        \texttt{ID} & Parameter & Default value & Scale \\
        \hline
        \texttt{1}&$m_d^{\overline{\rm MS}}$&\SI{0.00470(7)}{\gev}&\SI{2}{\gev}\\
        \texttt{2}&$m_u^{\overline{\rm MS}}$&\SI{0.00216(7)}{\gev}&\SI{2}{\gev}\\
        \texttt{3}&$m_s^{\overline{\rm MS}}$&\SI{0.0929(7)}{\gev}&\SI{2}{\gev}\\
        \texttt{4}&$m_c^{\overline{\rm MS}}$&\SI{1.2729(45)}{\gev}&$m_c$\\
        \texttt{11}&$m_e$&\SI{0.00051099895069(16)}{\gev}&--\\
        \texttt{13}&$m_\mu$&\SI{0.1056583755(23)}{\gev}&--\\
        \texttt{15}&$m_\tau$&\SI{1.77693(9)}{\gev}&--\\
        \texttt{24}&$M_{W}^{\rm pole}$&\SI{80.3625(77)}{\gev}&--\\
        \texttt{25}&$M_{h}^{\rm pole}$&\SI{125.13(11)}{\gev}&--\\
        \hline
    \end{tabular}
    \caption{Entries of the \texttt{MASS} block, from PDG \cite{ParticleDataGroup:2026aaa}.}
    \label{tab:MASS}
\end{table}

\paragraph{Block \texttt{VCKMIN}}
This block is defined by the SLHA convention. Default values are gathered in Table \ref{tab:VCKMIN}.

\renewcommand{\arraystretch}{1.3}
\begin{table}[!ht]
    \centering
    \begin{tabular}{cccc}
        \hline
        \texttt{ID} & Parameter & Default value \\
        \hline
        \texttt{1}&$\lambda$&\num{0.22517(68)}\\
        \texttt{2}&$A$&\num{0.826(17)}\\
        \texttt{3}&$\rho$&\num{0.1576(92)}\\
        \texttt{4}&$\eta$&\num{0.3556(71)}\\
        \hline
    \end{tabular}
    \caption{Entries of the \texttt{VCKMIN} block, from PDG \cite{ParticleDataGroup:2026aaa}.}
    \label{tab:VCKMIN}
\end{table}

\subsection{Flavour Parameters}
Here we only consider flavour parameters that are explicitly defined in the FLHA format \cite{Sakaki:2013bfa}. Flavour parameters (such as specific form factors, hadronic and leptonic moments\dots) which could technically be classified in the \texttt{FPARAM} block are instead stored in observable-specific custom blocks (see \ref{sec:obs_params}) for improved readability.

\paragraph{Block \texttt{FLIFE}}
This block contains particle lifetimes in seconds, indexed by the corresponding PDG Monte Carlo particle code. The default values used by \hi~are gathered in Table~\ref{tab:FLIFE}.
\begin{table}[!ht]
    \centering
    \begin{tabular}{ccc}
        \hline
        \texttt{ID} & Parameter & Default value \\
        \hline
        \texttt{130}&$\tau_{K_L^0}$&\SI{5.0990(21)e-8}{\second}\\
        \texttt{211}&$\tau_{\pi^+}$&\SI{2.6033(5)e-8}{\second}\\
        \texttt{310}&$\tau_{K_S^0}$&\SI{8.954(4)e-11}{\second}\\
        \texttt{321}&$\tau_{K^+}$&\SI{1.238(2)e-8}{\second}\\
        \texttt{323}&$\tau_{K^{*+}}$&\SI{1.280(2)e-23}{\second}\\
        \texttt{411}&$\tau_{D^+}$&\SI{1.033(5)e-12}{\second}\\
        \texttt{421}&$\tau_{D^0}$&\SI{4.103(10)e-13}{\second}\\
        \texttt{431}&$\tau_{D_s^+}$&\SI{5.012(22)e-13}{\second}\\
        \texttt{511}&$\tau_{B^0}$&\SI{1.517(4)e-12}{\second}\\
        \texttt{521}&$\tau_{B^+}$&\SI{1.638(4)e-12}{\second}\\
        \texttt{531}&$\tau_{B_s^0}$&\SI{1.527(11)e-12}{\second}\\
        \texttt{5122}&$\tau_{\Lambda_b^0}$&\SI{1.466(10)e-12}{\second}\\
        \hline
    \end{tabular}
    \caption{Entries of the \texttt{FLIFE} block, from PDG \cite{ParticleDataGroup:2026aaa}}
    \label{tab:FLIFE}
\end{table}

\paragraph{Block \texttt{FMASS}}
This block contains the masses of hadrons entering the flavour observables, indexed by their PDG Monte Carlo particle code. The defaults are expressed in GeV and are gathered in Table~\ref{tab:FMASS}.
\begin{table}[!ht]
    \centering
    \small
    \begin{tabular}{ccc}
        \hline
        \texttt{ID} & Parameter & Default value\\
        \hline
        \texttt{111}&$m_{\pi^0}$&\SI{0.1349768(5)}{\gev}\\
        \texttt{113}&$m_{\rho^0}$&\SI{0.76968(35)}{\gev}\\
        \texttt{211}&$m_{\pi^+}$&\SI{0.13957039(18)}{\gev}\\
        \texttt{311}&$m_{K^0}$&\SI{0.497611}{\gev}\\
        \texttt{313}&$m_{K^{*0}}$&\SI{0.89555(26)}{\gev}\\
        \texttt{321}&$m_{K^+}$&\SI{0.493677(13)}{\gev}\\
        \texttt{323}&$m_{K^{*+}}$&\SI{0.89167(26)}{\gev}\\
        \texttt{333}&$m_{\phi}$&\SI{1.01946(16)}{\gev}\\
        \texttt{411}&$m_{D^+}$&\SI{1.86966(5)}{\gev}\\
        \texttt{413}&$m_{D^{*+}}$&\SI{2.01026}{\gev}\\
        \texttt{421}&$m_{D^0}$&\SI{1.86484(5)}{\gev}\\
        \texttt{423}&$m_{D^{*0}}$&\SI{2.00685(5)}{\gev}\\
        \texttt{431}&$m_{D_s^+}$&\SI{1.96835(7)}{\gev}\\
        \texttt{443}&$m_{J/\psi}$&\SI{3.096900(6)}{\gev}\\
        \texttt{511}&$m_{B^0}$&\SI{5.27972(8)}{\gev}\\
        \texttt{521}&$m_{B^+}$&\SI{5.27941(7)}{\gev}\\
        \texttt{531}&$m_{B_s^0}$&\SI{5.36693(10)}{\gev}\\
        \texttt{3122}&$m_{\Lambda^0}$&\SI{1.115683(6)}{\gev}\\
        \texttt{5122}&$m_{\Lambda_b^0}$&\SI{5.61957(16)}{\gev}\\
        \texttt{100443}&$m_{\psi(2S)}$&\SI{3.686097(10)}{\gev}\\
        \hline
    \end{tabular}
    \caption{Entries of the \texttt{FMASS} block, from PDG \cite{ParticleDataGroup:2026aaa}}
    \label{tab:FMASS}
\end{table}

\paragraph{Block \texttt{FCONST}} 
This block contains the leptonic decay constants $f_P$ of pseudoscalar mesons, which are defined from the matrix element of the axial current between the meson state and the vacuum,
\begin{equation}
    \mel{0}{\bar q_1\gamma_\mu\gamma^5q_2}{P(p)}=-if_Pp_\mu.
\end{equation}
For vector mesons, the second index distinguishes longitudinal ($\parallel$) and transverse ($\perp$) decay constants. The transverse constants are scale dependent; the distributed defaults use $\mu=\SI{1}{\gev}$. The complete set of default values is gathered in Table~\ref{tab:FCONST}.
\begin{table}[!ht]
    \centering
    \begin{tabular}{ccccc}
        \hline
        \texttt{ID} & Parameter & Default value & Ref. & Scale \\
        \hline
        \texttt{211\_1}&$f_{\pi^+}$&\SI{0.1302(12)}{\gev}&\cite{FLAG:2024oxs}&\SI{1}{\gev}\\
        \texttt{311\_1}&$f_{K^0}$&\SI{0.1556(7)}{\gev}&\cite{FLAG:2024oxs}&\SI{1}{\gev}\\
        \texttt{321\_1}&$f_{K^+}$&\SI{0.1557(7)}{\gev}&\cite{FLAG:2024oxs}&\SI{1}{\gev}\\
        \texttt{313\_1}&$f_{K^{*0}}^{\parallel}$&\SI{0.204(7)}{\gev}&\cite{Ball:1998sk}&\SI{1}{\gev}\\
        \texttt{313\_2}&$f_{K^{*0}}^{\perp}$&\SI{0.159(6)}{\gev}&\cite{Ball:1998sk}&\SI{1}{\gev}\\
        \texttt{323\_1}&$f_{K^{*+}}^{\parallel}$&\SI{0.204(7)}{\gev}&\cite{Ball:1998sk}&\SI{1}{\gev}\\
        \texttt{323\_2}&$f_{K^{*+}}^{\perp}$&\SI{0.159(6)}{\gev}&\cite{Ball:1998sk}&\SI{1}{\gev}\\
        \texttt{333\_1}&$f_{\phi}^{\parallel}$&\SI{0.233(4)}{\gev}&\cite{Ball:1998sk}&\SI{1}{\gev}\\
        \texttt{333\_2}&$f_{\phi}^{\perp}$&\SI{0.191(4)}{\gev}&\cite{Ball:1998sk}&\SI{1}{\gev}\\
        \texttt{411\_1}&$f_{D^+}$&\SI{0.2126(7)}{\gev}&\cite{FLAG:2024oxs}&\SI{1}{\gev}\\
        \texttt{511\_1}&$f_{B^0}$&\SI{0.1900(13)}{\gev}&\cite{FLAG:2024oxs}&\SI{1}{\gev}\\
        \texttt{521\_1}&$f_{B^+}$&\SI{0.1900(13)}{\gev}&\cite{FLAG:2024oxs}&\SI{1}{\gev}\\
        \texttt{531\_1}&$f_{B_s^0}$&\SI{0.234(10)}{\gev}&\cite{FLAG:2024oxs}&\SI{1}{\gev}\\
        \hline
    \end{tabular}
    \caption{Entries of the \texttt{FCONST} block.}
    \label{tab:FCONST}
\end{table}

\paragraph{Block \texttt{FCONSTRATIO}} 
This block contains the ratio of leptonic decay constants of pseudoscalar mesons. Default values are gathered in Table \ref{tab:FCONSTRATIO}.

\begin{table}[!ht]
    \centering
    \begin{tabular}{ccccc}
        \hline
        \texttt{ID} & Parameter & Default value & Ref. & Scale \\
        \hline
        \texttt{321\_211\_1\_1}&$f_{K}/f_{\pi}$&\SI{1.1916(34)}{\gev}&\cite{FLAG:2024oxs}&\SI{1}{\gev}\\
        \hline
    \end{tabular}
    \caption{Entries of the \texttt{FCONSTRATIO} block.}
    \label{tab:FCONSTRATIO}
\end{table}

\subsection{Observable related parameters}\label{sec:obs_params}
\paragraph{Block \texttt{B\_Ks}} 
This block contains the parameters needed to compute observables related to the $B\to K^*(\to K\pi)\ell^+\ell^-$ decay. Due to the different values of the $B\to K^*$ form factors leading to phenomenologically significant differences in the final observables, \hi~ lets the user choose among several options gathered in the enum \texttt{BKstarllDecay::FF\_SRC}. The possible options are:
\begin{enumerate}[label=$\diamond$]
    \item \texttt{FF\_SRC::BSZ\_SR\_LAT}: LCSR + Lattice fit from \cite{Bharucha:2015bzk} (Table 15 p. 56)
    \item \texttt{FF\_SRC::BSZ\_SR}: LCSR only fit from \cite{Bharucha:2015bzk} (Table 14 p. 55)
    \item \texttt{FF\_SRC::GRvDV}: LCSR + Lattice fit from \cite{Gubernari:2023puw}
    \item \texttt{FF\_SRC::GKvD\_SR\_LAT}: LCSR + Lattice fit from \cite{Gubernari:2018wyi} 
    \item \texttt{FF\_SRC::GKvD\_SR}: LCSR only fit from \cite{Gubernari:2018wyi} 
    \item \texttt{FF\_SRC::HLMW}: Lattice fit from \cite{Horgan:2013hoa} and \cite{Horgan:2015vla} (Tables 5 and 6 p. 9)
\end{enumerate}
These parameters are stored within the block under the format \texttt{i\_j\_k} representing the series expansion parameter $\alpha_j^{(k)}$ where $i$ is the form factor choice id (in enum order), $j$ is the form factor id (in enum \texttt{BV::FF} order) and $k$ is the series expansion order. Default values are gathered in Table \ref{tab:B_Ks}.
\begin{table}[!ht]
    \centering
    \begin{tabular}{cccccc}
        \toprule
        \texttt{ID} & Parameter & Default value & Ref. & Scale & Comment \\
        \midrule
        \texttt{7\_1}&$a_1^\perp$&\num{0.04(7)}&\cite{Ball:2004rg}&$\SI{1}{\gev}$&--\\
        \texttt{7\_2}&$a_2^\perp$&\num{0.10(8)}&\cite{Ball:2004rg}&$\SI{1}{\gev}$&--\\
        \texttt{8\_1}&$a_1^\parallel$&\num{0.06(07)}&\cite{Ball:2004rg}&$\SI{1}{\gev}$&--\\
        \texttt{8\_2}&$a_2^\parallel$&\num{0.16(5)}&\cite{Ball:2004rg}&$\SI{1}{\gev}$&--\\
        \texttt{9}&$\zeta_3^A$&\num{0.032}&\cite{Ball:1998sk}&$\SI{1}{\gev}$&--\\
        \texttt{10}&$\zeta_3^V$&\num{0.013}&\cite{Ball:1998sk}&$\SI{1}{\gev}$&--\\
        \texttt{11}&$\omega_{1,0}^A$&\num{-2.1}&\cite{Ball:1998sk}&$\SI{1}{\gev}$&--\\
        \texttt{12\_1}&$\tilde\delta_+$&\num{0.16}&\cite{Ball:1998sk}&$\SI{1}{\gev}$&--\\
        \texttt{12\_2}&$\tilde\delta_-$&\num{-0.16}&\cite{Ball:1998sk}&$\SI{1}{\gev}$&--\\
        \texttt{13}&$\lambda_B$&\SI{0.46(11)}{\gev}&\cite{Ball:2006nr}&$\SI{1}{\gev}$&--\\
        \texttt{14}&$\Lambda_h$&\SI{0.5}{\gev}&--&--&Hadronic scale\\
        \texttt{15\_1}&$q^2_{\rm low}$&\SI{9}{\gev\squared}&--&--&Low-recoil boundary\\
        \texttt{15\_2}&$q^2_{\rm high}$&\SI{14}{\gev\squared}&--&--&High-$q^2$ boundary\\
        \texttt{16}&$T_1^{B\to K^*}(0)$&\num{0.312(32)}&--&$\SI{1}{\gev}$&Legacy normalisation input\\
        \texttt{17}&$\mu_0$&\SI{-1}{\gev}&--&--&Runtime sentinel\\
        \bottomrule
    \end{tabular}
    \caption{Selected legacy and QCD-factorisation entries of the \texttt{B\_Ks} block. The identifiers and numerical values match the \releaseversion~ default asset. The standalone $T_1(0)$ entry is retained as an unreferenced legacy normalisation input in the released database; the form-factor parametrisations used for semileptonic predictions are documented separately below.}
    \label{tab:B_Ks}
\end{table}
\footnotetext{When this parameter is set to $-1$, then $\mu_0=\mu_b$ is enforced at runtime.}

\paragraph{Block \texttt{B\_ll}} 
This block contains the parameters needed to compute observables related to $B\to \ell^+\ell^-$ decays. Default values are gathered in Table \ref{tab:B_ll}.
\begin{table}[!ht]
    \centering
    \begin{tabular}{cccc}
        \hline
        \texttt{ID} & Parameter & Default value & Ref. \\
        \hline
        \texttt{1}&$y_s$&\num{0.0635(14)}&\cite{ParticleDataGroup:2024cfk}\\
        \texttt{2}&$\eta_{BBS}$&\num{0.995(5)}&\cite{czaja2024currentstatusstandardmodel}\\
        \hline
    \end{tabular}
    \caption{Entries of the \texttt{B\_ll} block.}
    \label{tab:B_ll}
\end{table}

\paragraph{Block \texttt{B\_Xs}} 
This block contains the parameters needed to compute observables related to the $B\to X_s\gamma$ inclusive decay. Default values are gathered in Table \ref{tab:B_Xs}.
\begin{table}[!ht]
    \centering
    \begin{tabular}{cccccc}
        \hline
        \texttt{ID} & Parameter & Default value & Ref. & Comment \\
        \hline
        \texttt{1}&$E_0$&\SI{1.6}{\gev}&--&--\\
        \texttt{2}&$\mathrm{BR}(B\to X_c e\bar\nu)_{\rm exp.}$&\num{0.1065(16)}&\cite{Gambino:2013rza}&--\\
        \texttt{3}&$\mu_G^2$&\num{0.336(64)}&\cite{Gambino:2013rza}&--\\
        \texttt{4}&$\rho_D^3$&\num{0.153(45)}&\cite{Gambino:2013rza}&--\\
        \texttt{5}&$\rho_{LS}^3$&\num{-0.145(98)}&\cite{Gambino:2013rza}&--\\
        \texttt{6}&$\lambda_2$&\SI{0.120(18)}{\gev\squared}&\cite{ParticleDataGroup:2024cfk,Misiak:2006ab}&--\\
        \texttt{7}&$\mu_c$&\SI{2.45(1.55)}{\gev}&--&--\\
        \texttt{8}&$z_0$&$10^{10}$&--&For interpolation\\
        \texttt{9}&$z_1$&$10^{20}$&--&For interpolation\\
        \hline
    \end{tabular}
    \caption{Entries of the \texttt{B\_Xs} block.}
    \label{tab:B_Xs}
\end{table}

\paragraph{Block \texttt{B\_Dlnu}} 
This block contains the parameters needed to compute observables related to $B\to D$ decays. Default values are gathered in Table \ref{tab:B_Dlnu}.
\begin{table}[!ht]
    \centering
    \begin{tabular}{cccccc}
        \hline
        \texttt{ID} & Parameter & Default value & Ref. \\
        \hline
        \texttt{1}&$V_1(1)$&\num{1.074(24)}&\cite{Sakaki:2013bfa}\\
        \texttt{2}&$\rho_D^2$&\num{1.186(54)}&\cite{Sakaki:2013bfa}\\
        \texttt{3}&$\Delta(w)$&\num{1(1)}&\cite{Sakaki:2013bfa}\\
        \hline
    \end{tabular}
    \caption{Entries of the \texttt{B\_Dlnu} block.}
    \label{tab:B_Dlnu}
\end{table}

\paragraph{Block \texttt{B\_Dslnu}} 
This block contains the parameters needed to compute observables related to $B\to D^*$ decays. Default values are gathered in Table \ref{tab:B_Dslnu}.
\begin{table}[!ht]
    \centering
    \begin{tabular}{cccccc}
        \hline
        \texttt{ID} & Parameter & Default value & Ref. \\
        \hline
        \texttt{1}&$h_{A_1}(1)$&\num{0.921(17)}&\cite{Sakaki:2013bfa}\\
        \texttt{2}&$\rho_{D^*}^2$&\num{1.214(26)}&\cite{Sakaki:2013bfa}\\
        \texttt{3}&$R_1(1)$&\num{1.403(33)}&\cite{Sakaki:2013bfa}\\
        \texttt{4}&$R_2(1)$&\num{0.864(20)}&\cite{Sakaki:2013bfa}\\
        \hline
    \end{tabular}
    \caption{Entries of the \texttt{B\_Dslnu} block.}
    \label{tab:B_Dslnu}
\end{table}

\paragraph{Block \texttt{B\_K}} 
This block contains the parameters needed to compute observables related to $B\to K$ decays. As for the $B\to K^*\ell\ell$ decay there are several form factor choice implemented in \hi{} which are stored in the \texttt{BKllDecay::FF\_SRC} enumeration as follows: 
\begin{enumerate}[label=$\diamond$]
    \item \texttt{BP\_FF\_Src::AS}: LCSR + lattice fit for the
    \(B\to K\) form factors from \cite{Altmannshofer:2014rta}
    (Appendix~A, Eqs.~(51)--(53)).

    \item \texttt{BP\_FF\_Src::GRvDV}: Dispersive fit combining LCSR
    and lattice constraints from \cite{Gubernari:2023puw}.

    \item \texttt{BP\_FF\_Src::GKvD\_SR\_LAT}: Fit combining the LCSR
    results of \cite{Gubernari:2018wyi} with lattice constraints.

    \item \texttt{BP\_FF\_Src::GKvD\_SR}: LCSR-only fit based on
    \cite{Gubernari:2018wyi}.

    \item \texttt{BP\_FF\_Src::FLAG24}: Lattice average from the
    FLAG Review 2024 \cite{FLAG:2024oxs}.

    \item \texttt{BP\_FF\_Src::HPQCD22}: Fully relativistic lattice
    fit over the full physical \(q^2\) region from the HPQCD
    collaboration \cite{Parrott:2022rgu}.
\end{enumerate}
Default values are gathered in Table \ref{tab:B_K}.
 
\begin{table}[!ht]
    \centering
    \begin{tabular}{ccccccc}
        \hline
        \texttt{ID} & Parameter & Default value & Ref. & Scale & Comment \\
        \hline
        \texttt{8\_1}&$a_1$&\num{0.06(3)}&\cite{Ball:2006nr}&\SI{1}{\gev}\\
        \texttt{8\_2}&$a_2$&\num{0.25(15)}&\cite{Ball:2006nr}&\SI{1}{\gev}\\
        \texttt{13}&$\lambda_B$&\SI{0.46(11)}{\gev}&\cite{Ball:2006nr}&\SI{1}{\gev}&--\\
        \texttt{14}&$\Lambda_h$&\SI{0.5}{\gev}&--&--&Hadronic scale\\
        \texttt{15\_1}&$q^2_{\rm low}$&\SI{9}{\gev\squared}&--&--&Low-recoil boundary\\
        \texttt{15\_2}&$q^2_{\rm high}$&\SI{14}{\gev\squared}&--&--&High-$q^2$ boundary\\
        \texttt{16}&$L_\chi$&1.304&\cite{Parrott:2022rgu}&--\\
        \hline
    \end{tabular}
    \caption{Entries of the \texttt{B\_K} block.}
    \label{tab:B_K}
\end{table}

\paragraph{Block \texttt{B\_Xsll}} 
This block contains the parameters needed to compute observables related to $B\to X_s\ell\ell$ inclusive decay. Default values are gathered in Table \ref{tab:B_Xsll}.
\begin{table}[!ht]
    \centering
    \begin{tabular}{cccccc}
        \hline
        \texttt{ID} & Parameter & Default value & Ref. \\
        \hline
        \texttt{1}&$\lambda_1$&\SI{-0.15(10)}{\gev\squared}&\cite{Proceedings:2003xqt}\\
        \texttt{2}&$\rho_1$&\SI{0.06(6)}{\gev\cubed}&\cite{Proceedings:2003xqt}\\
        \hline
    \end{tabular}
    \caption{Entries of the \texttt{B\_Xsll} block.}
    \label{tab:B_Xsll}
\end{table}

\paragraph{Block \texttt{K\_ll}} 
This block contains the parameters needed to compute observables related to $K_{L,S}^0\to\ell\ell$ decay. Default values are gathered in Table \ref{tab:K_ll}.
\begin{table}[!ht]
    \centering
    \begin{tabular}{cccc}
        \hline
        \texttt{ID} & Parameter & Default value & Ref. \\
        \hline
        \texttt{1}
        & $\mathrm{BR}(K_L^0\to\gamma\gamma)_{\rm exp.}$
        & \num{5.47(4)e-4}
        & \cite{ParticleDataGroup:2026aaa}
        \\

        \texttt{2}
        & $\mathrm{BR}(K_S^0\to\gamma\gamma)_{\rm exp.}$
        & \num{2.63(17)e-6}
        & \cite{ParticleDataGroup:2026aaa}
        \\

        \texttt{3}
        & $\alpha_{\rm exp.}$
        & \num{-1.611(44)}
        & \cite{Isidori:2003ts}
        \\

        \texttt{4}
        & $\Delta_\Lambda$
        & \num{0(1)}
        & \cite{Isidori:2003ts}
        \\
        \hline
    \end{tabular}
    \caption{Entries of the \texttt{K\_ll} block.}
    \label{tab:K_ll}
\end{table}

\paragraph{Block \texttt{K\_lnu}} 
This block contains the parameters needed to compute observables related to $K\to\ell\nu$ decay. Default values are gathered in Table \ref{tab:K_lnu}.
\begin{table}[!ht]
    \centering
    \begin{tabular}{cccccc}
        \hline
        \texttt{ID} & Parameter & Default value & Ref. \\
        \hline
        \texttt{1}&$\delta_{\rm em}$&\num{70(35)e-4}&\cite{Antonelli_2010}\\
        \hline
    \end{tabular}
    \caption{Entries of the \texttt{K\_lnu} block.}
    \label{tab:K_lnu}
\end{table}

\paragraph{Block \texttt{K\_pi}} 
This block contains the parameters needed to compute observables related to $K\to\pi\nu\nu$ decay. Default values are gathered in Table \ref{tab:K_pi}.
\begin{table}[!ht]
    \centering
    \begin{tabular}{cccccc}
        \hline
        \texttt{ID} & Parameter & Default value & Ref. \\
        \hline
        \texttt{1}&$\kappa_{L}$&\num{2.231(13)}&\cite{Brod_2011}\\
        \texttt{2}&$\kappa_{+}$&\num{0.5173(25)}&\cite{Brod_2011}\\
        \texttt{3}&$\Delta_{\rm em}$&\num{-0.003}&\cite{Mescia_2007}\\
        \hline
    \end{tabular}
    \caption{Entries of the \texttt{K\_pi} block.}
    \label{tab:K_pi}
\end{table}

\paragraph{Block \texttt{B\_phi}} 
This block contains the parameters needed to compute observables related to the $B_s\to \phi\ell^+\ell^-$ decay. The form factor choice is the same as for $B\to K^*\ell\ell$ decays, except for those from GKvD. Default values are gathered in Table \ref{tab:B_phi}. 
\begin{table}[!ht]
    \centering
    \begin{tabular}{cccccc}
        \toprule
        \texttt{ID} & Parameter & Default value & Ref. & Scale & Comment \\
        \midrule
        \texttt{7\_1}&$a_1^\perp$&\num{0.0}&\cite{Ball:2004rg}&$\SI{1}{\gev}$&--\\
        \texttt{7\_2}&$a_2^\perp$&\num{0.14(7)}&\cite{Ball:2004rg}&$\SI{1}{\gev}$&--\\
        \texttt{8\_1}&$a_1^\parallel$&\num{0.0}&\cite{Ball:2004rg}&$\SI{1}{\gev}$&--\\
        \texttt{8\_2}&$a_2^\parallel$&\num{0.23(8)}&\cite{Ball:2004rg}&$\SI{1}{\gev}$&--\\
        \texttt{9}&$\zeta_3^A$&\num{0.032}&\cite{Ball:1998sk}&$\SI{1}{\gev}$&--\\
        \texttt{10}&$\zeta_3^V$&\num{0.013}&\cite{Ball:1998sk}&$\SI{1}{\gev}$&--\\
        \texttt{11}&$\omega_{1,0}^A$&\num{-2.1}&\cite{Ball:1998sk}&$\SI{1}{\gev}$&--\\
        \texttt{12\_1}&$\tilde\delta_+$&\num{0.33}&\cite{Ball:1998sk}&$\SI{1}{\gev}$&--\\
        \texttt{12\_2}&$\tilde\delta_-$&\num{0.0}&\cite{Ball:1998sk}&$\SI{1}{\gev}$&--\\
        \texttt{13}&$\lambda_B$&\SI{0.46(11)}{\gev}&\cite{Ball:2006nr}&$\SI{1}{\gev}$&--\\
        \texttt{14}&$\Lambda_h$&\SI{0.5}{\gev}&--&--&Hadronic scale\\
        \texttt{15\_1}&$q^2_{\rm low}$&\SI{9}{\gev\squared}&--&--&Low-recoil boundary\\
        \texttt{15\_2}&$q^2_{\rm high}$&\SI{14}{\gev\squared}&--&--&High-$q^2$ boundary\\
        \bottomrule
    \end{tabular}
    \caption{Selected entries of the \texttt{B\_phi} block. The form-factor parametrisations used for semileptonic predictions are documented separately below.}
    \label{tab:B_phi}
\end{table}

\paragraph{Block \texttt{Lb\_L}} 
This block contains the parameters needed to compute observables related to the $\Lambda_b\to\Lambda\ell\ell$ decay. There is currently only one form factor choice taken from \cite{Detmold:2016pkz} and follow the same naming conventions as the other decays. Default values are gathered in Table \ref{tab:Lb_L}.
\begin{table}[!ht]
    \centering
    \begin{tabular}{cccccc}
        \hline
        \texttt{ID} & Parameter & Default value & Ref. \\
        \hline
        \texttt{1}&$\alpha_{\Lambda}$&\num{0.642(13)}&\cite{Detmold:2016pkz}\\
        \hline
    \end{tabular}
    \caption{Entries of the \texttt{Lb\_L} block.}
    \label{tab:Lb_L}
\end{table}

\section{Observable catalogue by decay family}\label{app:observable_catalogue}

This appendix gives a decay-by-decay view (see Tables \ref{tab:observable_catalogue_b_rare}, \ref{tab:observable_catalogue_charged_current} and \ref{tab:observable_catalogue_kaon_mixing}) of the native observable coverage summarised in Table~\ref{tab:decays}.  The entries are grouped by physics channel rather than by internal C++ class name.  Canonical mapper names are intentionally not exhaustively listed here because the full list is generated by the observable registry and distributed with the code.

\begin{table}[!htbp]
\centering
\footnotesize
\renewcommand{\arraystretch}{1.15}
\begin{tabularx}{\textwidth}{@{}p{0.23\textwidth}p{0.28\textwidth}X@{}}
\toprule
\textbf{Decay family} & \textbf{Internal family} & \textbf{Representative observables} \\
\midrule
$B\to X_s\gamma$ & \texttt{B\_\_Xs} &
Inclusive radiative branching fraction and related radiative observables. \\
$B\to X_s\ell^+\ell^-$ & \texttt{B\_\_Xs\_ll} &
Inclusive rare-decay branching fractions and forward--backward asymmetries for $e$, $\mu$ and $\tau$ channels. \\
$B\to K^*\gamma$ & \texttt{B\_\_Kstar\_gamma} &
Exclusive radiative branching fractions and isospin asymmetries for charged and neutral modes. \\
$B\to K^*\ell^+\ell^-$ & \texttt{B\_\_Kstar\_l\_l} &
Differential rates, $A_{FB}$, $F_L$, $F_T$, $A_T^{(i)}$, $H_T^{(i)}$, $P_i^{(\prime)}$, $S_i$ and CP-asymmetry observables. \\
$B\to K\ell^+\ell^-$ & \texttt{B\_\_K\_l\_l} &
Differential rates, branching fractions, $A_{FB}$, $F_H$ and lepton-universality ratios. \\
$B_s\to\phi\ell^+\ell^-$ & \texttt{Bs\_\_phi\_l\_l} &
Differential rates, angular observables, optimised observables and CP-sensitive combinations. \\
$\Lambda_b\to\Lambda\ell^+\ell^-$ & \shortstack[l]{\texttt{Lambda\_b\_\_}\\\texttt{Lambda\_l\_l}} &
Differential rates, longitudinal/transverse polarisation fractions and lepton/hadron forward--backward asymmetries. \\
\bottomrule
\end{tabularx}
\caption{Native rare and radiative $B$-sector observables.}
\label{tab:observable_catalogue_b_rare}
\end{table}

\begin{table}[!ht]
\centering
\footnotesize
\renewcommand{\arraystretch}{1.15}
\begin{tabularx}{\textwidth}{@{}p{0.23\textwidth}p{0.28\textwidth}X@{}}
\toprule
\textbf{Decay family} & \textbf{Internal family} & \textbf{Representative observables} \\
\midrule
$B_{s,d}\to\ell^+\ell^-$ & \texttt{B\_\_l\_l} &
Leptonic branching fractions, untagged $B_s$ modes and related lifetime-sensitive quantities. \\
$B_u\to\ell\nu$ & \texttt{B\_\_l\_nu} &
Charged-$B$ leptonic branching fractions and charged-current ratios. \\
$B\to D\ell\nu$ & \texttt{B\_\_D\_l\_nu} &
Branching fractions, $R(D)$, forward--backward asymmetry and charged-lepton polarisation. \\
$B\to D^*\ell\nu$ & \shortstack[l]{\texttt{B\_\_Dstar}\\\texttt{\_l\_nu}} &
Branching fractions, $R(D^*)$, forward--backward asymmetry, $\tau$ polarisation and $D^*$ polarisation. \\
$D\to\ell\nu$ & \texttt{D\_\_l\_nu} &
Charm leptonic branching fractions. \\
$D_s\to\ell\nu$ & \texttt{Ds\_\_l\_nu} &
$D_s\to\mu\nu$ and $D_s\to\tau\nu$ branching fractions. \\
\bottomrule
\end{tabularx}
\caption{Native leptonic and charged-current observables.}
\label{tab:observable_catalogue_charged_current}
\end{table}

\begin{table}[!ht]
\centering
\footnotesize
\renewcommand{\arraystretch}{1.15}
\begin{tabularx}{\textwidth}{@{}p{0.23\textwidth}p{0.28\textwidth}X@{}}
\toprule
\textbf{Decay family} & \textbf{Internal family} & \textbf{Representative observables} \\
\midrule
$K\to\pi\nu\bar\nu$ & \texttt{K\_\_pi\_nu\_nu} &
$K^+\to\pi^+\nu\bar\nu$ and $K_L\to\pi^0\nu\bar\nu$ branching fractions. \\
$K_{L,S}\to\mu^+\mu^-$ & \texttt{K\_\_l\_l} &
Rare kaon dimuon branching fractions and short-/long-distance components. \\
$K\to\ell\nu$ & \texttt{K\_\_l\_nu} &
Leptonic kaon ratios such as $R_{\mu 23}$. \\
Neutral-meson mixing & \texttt{M0\_Mix} &
$\Delta M_{B_d}$, $\Delta M_{B_s}$, $\Delta M_K$, $|\epsilon_K|$, CP phases and semileptonic asymmetries. \\
Custom observable layer & runtime registry &
User-defined observables attached to native or user-defined decay families. \\
\bottomrule
\end{tabularx}
\caption{Native kaon, mixing and extension-layer observables.}
\label{tab:observable_catalogue_kaon_mixing}
\end{table}

\section{Leptonic decays of charged pseudoscalar mesons}\label{app:plnu}
The purely leptonic decays of pseudoscalar mesons $P^\pm\to\ell^\pm\nu$ happen in the SM through a virtual $W$ boson. In extensions of the SM with two Higgs doublets, charged scalars interfere destructively with the SM. The most general effective Hamiltonian for such decays (at dimension 6) reads:
\begin{equation}
    \mathcal{H}_{CC}=\frac{4G_F}{\sqrt{2}}V_{q_uq_d}\sum C_k\mathcal{O}_k+\mathrm{h.c.},
\end{equation}
where $V_{q_uq_d}$ is a CKM matrix element, and the 5 independent operators read:
\begin{gather*}
    \mathcal{O}_{V_1}=\qty(\bar q_u\gamma^\mu P_Lq_d)\qty(\bar\ell\gamma_\mu P_L\nu),\\
    \mathcal{O}_{V_2}=\qty(\bar q_u\gamma^\mu P_Rq_d)\qty(\bar\ell\gamma_\mu P_L\nu),\\
    \mathcal{O}_{S_1}=\qty(\bar q_uP_Lq_d)\qty(\bar\ell P_L\nu)\numberthis{},\\
    \mathcal{O}_{S_2}=\qty(\bar q_uP_Rq_d)\qty(\bar\ell P_L\nu),\\
    \mathcal{O}_{T}=\qty(\bar q_u\sigma^{\mu\nu} P_Lq_d)\qty(\bar\ell\sigma^{\mu\nu} P_L\nu).
\end{gather*}

Using then the definition of the meson's decay constant and the quark EoM,
\begin{equation}
    \mel{0}{\bar q_u\gamma^\mu\gamma^5 q_d}{\bar P}=ip^\mu f_P\qc \mel{0}{\bar q_u\gamma^5 q_d}{\bar P}=-i\frac{m_P^2}{m_{q_u}+m_{q_d}}f_P,
\end{equation}
\newpage

while the other currents do not contribute. The complete matrix elements for the effective operators then read:
\begin{gather*}
    \mel{\ell\bar\nu}{\mathcal{O}_{V_{1,2}}}{\bar P}=\mp\frac{i}{2}f_P\qty(\bar\ell\fsl{p}P_L\nu),\\
    \mel{\ell\bar\nu}{\mathcal{O}_{S_{1,2}}}{\bar P}=\pm\frac{i}{2}\frac{m_P^2}{m_{q_u}+m_{q_d}}f_P\qty(\bar\ell P_L\nu),\numberthis{}\\
    \mel{\ell\bar\nu}{\mathcal{O}_{T}}{\bar P}=0.
\end{gather*}
The decaying pseudoscalar meson has spin 0, and the outgoing (anti-)neutrino must be (right) left-handed, thus the outgoing (anti-)lepton must have the same chirality. Using the chirality eigenstates of Dirac spinors, and the CoM kinematics where $P$ is at rest and $\ell,\nu$ are back-to-back, the leptonic part of the matrix elements evaluates to:
\begin{equation}
    \bar\ell\fsl{p}P_L\nu=m_\ell m_P\sqrt{1-r_\ell}\qc \bar\ell P_L\nu=m_P\sqrt{1-r_\ell},
\end{equation}
where $r_\ell=m_\ell^2/m_P^2$. Further defining $C_A=C_{V_2}-C_{V_1}$ and $C_P=C_{S_2}-C_{S_1}$, the squared amplitude reads:
\begin{equation}
    \qty|\mathcal{M}|^2=2G_F^2\qty|V_{q_uq_d}|^2f_P^2m_P^2(1-r_\ell)\qty|m_\ell C_A-\frac{m_P^2}{m_{q_u}+m_{q_d}}C_P|^2.
\end{equation}
And finally, using
\begin{equation}
    \Gamma=\frac{\beta}{16\pi m_P}\qty|\mathcal{M}|^2\qc\beta=1-r_\ell,
\end{equation}
the branching ratio reads:
\begin{equation}
    \mathcal{B}(P\to\ell\nu)=\frac{G_F^2}{8\pi}|V_{q_uq_d}|^2f_P^2m_P\tau_P(1-r_\ell)^2\qty|m_\ell C_A-\frac{m_P^2}{m_{q_u}+m_{q_d}}C_P|^2 .
\end{equation}

\section{Extended validation tables}\label{app:extended_validation_tables}

This appendix collects the complete Wilson-coefficient validation tables.  The \spi--\hi~comparison is split into SM, THDM and SUSY benchmarks (see Tables \ref{tab:comparison_superiso_full_sm}, \ref{tab:comparison_superiso_full_thdm} and \ref{tab:comparison_superiso_full_susy}). The native-\hi/\mty~comparison is split into SM and type-II THDM BSM contributions at the matching scale.

\begin{table}[!ht]
\centering
\setlength{\tabcolsep}{4pt}
\renewcommand{\arraystretch}{1.1}
\begin{adjustbox}{max width=\textwidth}
\begin{tabular}{lSSS}
\hline
\textbf{coefs} & \textbf{SI} & \textbf{HI} & \textbf{\%} \\
\hline
\multicolumn{4}{c}{\(\mu_W=\mathcal{O}(M_W)\)} \\
\hline
\multicolumn{4}{c}{\textbf{SM}} \\
\hline
C1 & 0.1583963453 & 0.1583963451 & 0 \\
C2 & 1.001843983 & 1.001843983 & 0 \\
C3 & -0.0002931687782 & -0.0002931687782 & 0 \\
C4 & -0.003248811761 & -0.003247432028 & 0.042478 \\
C5 & 3.386474459e-05 & 3.386879571e-05 & 0.011962 \\
C6 & 6.349639566e-05 & 6.349639566e-05 & 0 \\
C7 & -0.2107427536 & -0.2106774232 & 0.031005 \\
C8 & -0.1151167202 & -0.1150915551 & 0.021863 \\
C9 & 1.930302269 & 1.92962599 & 0.035041 \\
C10 & -4.171695742 & -4.168626236 & 0.073606 \\
CQ1 & -0.00026039 & -0.000260394 & 0.001536 \\
CQ2 & 0.000485819217 & 0.000484792 & 0.000413 \\
CP7 & -0.00423707 & -0.00423707 & 0 \\
CP8 & -0.00221976 & -0.00221976 & 0 \\
CP9 & 0 & 0 & 0 \\
CP10 & 0 & 0 & 0 \\
CPQ1 & 0 & 0 & 0 \\
CPQ2 & 0 & 0 & 0 \\
\hline
\end{tabular}
\end{adjustbox}
\caption{Complete comparison between \spi~(SI) and \hi~(HI) for the SM benchmark at matching scale.}
\label{tab:comparison_superiso_full_sm}
\end{table}

\begin{table}[!ht]
\centering
\setlength{\tabcolsep}{4pt}
\renewcommand{\arraystretch}{1.1}
\begin{adjustbox}{max width=\textwidth}
\begin{tabular}{lSSS}
\hline
\textbf{coefs} & \textbf{SI} & \textbf{HI} & \textbf{\%} \\
\hline
\multicolumn{4}{c}{\(\mu_W=\mathcal{O}(M_W)\)} \\
\hline
\multicolumn{4}{c}{\textbf{THDM}} \\
\hline
C1 & 0 & 0 & 0 \\
C2 & 0 & 0 & 0 \\
C3 & -3.723792155e-08 & -3.723792155e-08 & 0 \\
C4 & 9.296889404e-06 & 9.296896376e-06 & 0.000075 \\
C5 & 1.400574929e-08 & 1.400577667e-08 & 0.000195 \\
C6 & 2.626083695e-08 & 2.626083695e-08 & 0 \\
C7 & 0.0008007874546 & 0.0008004786918 & 0.038565 \\
C8 & 0.0008704866784 & 0.0008704864666 & 0.000024 \\
C9 & -0.0002778553141 & -0.0002778544545 & 0.000309 \\
C10 & -0.004887954445 & -0.004887955214 & 0.000016 \\
CQ1 & -0.00038257 & -0.000382573 & 0.000784 \\
CQ2 & 6.37e-06 & 6.3716e-06 & 0.025115 \\
CP7 & 0 & 0 & 0 \\
CP8 & 0 & 0 & 0 \\
CP9 & 0 & 0 & 0 \\
CP10 & 0 & 0 & 0 \\
CPQ1 & 0 & 0 & 0 \\
CPQ2 & 0 & 0 & 0 \\
\hline
\end{tabular}
\end{adjustbox}
\caption{Complete comparison between \spi~(SI) and \hi~(HI) for the THDM benchmark at matching scale.}
\label{tab:comparison_superiso_full_thdm}
\end{table}

\begin{table}[!ht]
\centering
\setlength{\tabcolsep}{4pt}
\renewcommand{\arraystretch}{1.1}
\begin{adjustbox}{max width=\textwidth}
\begin{tabular}{lSSS}
\hline
\textbf{coefs} & \textbf{SI} & \textbf{HI} & \textbf{\%} \\
\hline
\multicolumn{4}{c}{\(\mu_W=\mathcal{O}(M_W)\)} \\
\hline
\multicolumn{4}{c}{\textbf{SUSY}} \\
\hline
C1 & -5.594240306e-07 & -5.594240306e-07 & 0 \\
C2 & 0 & 0 & 0 \\
C3 & 4.9344336e-07 & 4.934424585e-07 & 0.000183 \\
C4 & 4.003675142e-06 & 4.003666186e-06 & 0.000224 \\
C5 & -4.44468528e-08 & -4.444694295e-08 & 0.000203 \\
C6 & -8.333807436e-08 & -8.333798422e-08 & 0.000108 \\
C7 & 0.0008455493107 & 0.0008451479372 & 0.047480 \\
C8 & -0.0101108152 & -0.010110883 & 0.000671 \\
C9 & 0.00588099578 & 0.005879457677 & 0.026157 \\
C10 & 0.005718707991 & 0.00571869065 & 0.000303 \\
CQ1 & -0.0004304081933 & -0.0004303745802 & 0.007810 \\
CQ2 & 0.0001911053555 & 0.0001910728092 & 0.017032 \\
CP7 & 0.000219408 & 0.000219408 & 0 \\
CP8 & 1.0434e-05 & 1.04337e-05 & 0.002875 \\
CP9 & -2.968e-06 & -2.96754e-06 & 0.015500 \\
CP10 & 7.0043e-05 & 7.0043e-05 & 0 \\
CPQ1 & -6.9317e-06 & -6.93174e-06 & 0.000577 \\
CPQ2 & -6.9315e-06 & -6.93149e-06 & 0.000144 \\
\hline
\end{tabular}
\end{adjustbox}
\caption{Complete comparison between \spi~(SI) and \hi~(HI) for the SUSY benchmark at matching scale.}
\label{tab:comparison_superiso_full_susy}
\end{table}

Tables~\ref{tab:comparison_marty_full_sm} and~\ref{tab:comparison_marty_full_bsm} show the displayed native-\hi/\mty~LO export at the matching scale for the type-II THDM benchmark.  Coefficients that vanish in both calculations are omitted.  The printed validation subset keeps the standard \(B\)-sector coefficients, the non-zero primed coefficients and the muonic scalar coefficients \(C_{Q_1}^\mu\) and \(C_{Q_2}^\mu\).  Electron and tau scalar entries, as well as the \(C^\prime_Q\) family, are omitted from the paper tables for compactness.
\newpage

\begin{table}[!ht]
\centering
\small
\setlength{\tabcolsep}{4pt}
\renewcommand{\arraystretch}{1.08}
\begin{adjustbox}{max width=0.9\textwidth}
\begin{tabular}{llccc}
\hline
\textbf{Group} & \textbf{Coefficient} & \textbf{HI} & \textbf{\mty} & \(\boldsymbol{\delta_{\rm rel}}\) \\
\hline
\(B\) & \(C_{2}\) & \num{1} & \num{1.000064} & \num{6.4e-05} \\
\(B\) & \(C_{7}\) & \num{-0.19411475} & \num{-0.1942087} & \num{0.000484} \\
\(B\) & \(C_{8}\) & \num{-0.096902814} & \num{-0.09697558} & \num{0.000751} \\
\(B\) & \(C_{9}\) & \num{2.0076248} & \num{2.0076248} & \num{0} \\
\(B\) & \(C_{10}\) & \num{-4.4637361} & \(\num{-4.479714}+\num{0.08860065}\,i\) & \num{0.0202} \\
\(B^\prime\) & \(C^\prime_{7}\) & \num{-0.0066274482} & \num{-0.00653486} & \num{0.014} \\
\(B^\prime\) & \(C^\prime_{8}\) & \num{-0.0033084471} & \num{-0.003262563} & \num{0.0139} \\
\(B_{\rm S}\) & \(C_{Q_1}^{\mu}\) & \num{-0.00036956518} & \num{-0.0003589897} & \num{0.0289} \\
\(B_{\rm S}\) & \(C_{Q_2}^{\mu}\) & \num{0.00033655824} & \num{0.0003354648} & \num{0.00464} \\
\hline
\end{tabular}
\end{adjustbox}
\caption{Complete displayed native-\hi/\mty~comparison for the SM contribution at matching scale.  The comparison is restricted to LO, the order generated by the \mty~interface in this workflow.  The last column gives the relative difference defined in Eq.~\ref{eq:marty_relative_difference}.}
\label{tab:comparison_marty_full_sm}
\end{table}

\begin{table}[!ht]
\centering
\small
\setlength{\tabcolsep}{4pt}
\renewcommand{\arraystretch}{1.08}
\begin{adjustbox}{max width=0.9\textwidth}
\begin{tabular}{llccc}
\hline
\textbf{Group} & \textbf{Coefficient} & \textbf{HI} & \textbf{\mty} & \(\boldsymbol{\delta_{\rm rel}}\) \\
\hline
\(B\) & \(C_{7}\) & \num{-0.099068547} & \num{-0.0990574} & \num{0.000113} \\
\(B\) & \(C_{8}\) & \num{-0.099275045} & \num{-0.09925882} & \num{0.000163} \\
\(B\) & \(C_{9}\) & \num{0.00012406279} & \num{0.0001228477} & \num{0.00979} \\
\(B\) & \(C_{10}\) & \num{-0.0065801945} & \num{-0.006560991} & \num{0.00292} \\
\(B^\prime\) & \(C^\prime_{7}\) & \num{-0.0033823893} & \num{-0.003327419} & \num{0.0163} \\
\(B^\prime\) & \(C^\prime_{8}\) & \num{-0.0033894395} & \num{-0.003333466} & \num{0.0165} \\
\(B^\prime\) & \(C^\prime_{9}\) & \num{-7.3827472e-05} & \num{-7.365517e-05} & \num{0.00233} \\
\(B^\prime\) & \(C^\prime_{10}\) & \num{0.00055986823} & \num{0.000557905} & \num{0.00351} \\
\(B_{\rm S}\) & \(C_{Q_1}^{\mu}\) & \num{-0.0022734188} & \(\num{-0.0018731}-\num{0.000322062}\,i\) & \num{0.226} \\
\(B_{\rm S}\) & \(C_{Q_2}^{\mu}\) & \num{0.0020599269} & \(\num{0.001624921}+\num{0.0003148968}\,i\) & \num{0.261} \\
\hline
\end{tabular}
\end{adjustbox}
\caption{Complete displayed native-\hi/\mty~comparison for the genuine type-II THDM BSM contribution at matching scale.  The comparison is restricted to LO, the order generated by the \mty~interface in this workflow.  The last column gives the relative difference defined in Eq.~\ref{eq:marty_relative_difference}.}
\label{tab:comparison_marty_full_bsm}
\end{table}

\section{Efficient computation of the profiled likelihood}
For a given set of observables, there are usually many nuisance parameters (typically $\mathcal{O}(100)$) that we need to optimise in order to compute the profiled likelihood function $\hat\ell(p)$. Even the best algorithms cannot solve such high-dimensional optimisation problems efficiently enough, especially as this step would be performed thousands of times to compute the confidence contours. We need to use some additional information that we may have on the nuisances to reduce the dimensionality of this problem. One way we can do this is to separate the nuisances in two groups, the ``well-behaved'' $\eta_W$ and the ``ill-behaved'' $\eta_I$. A nuisance parameter $\eta_a$ is``well-behaved'' if its marginal distribution is Gaussian and the model function $f(p,\eta)$ is linear in $\eta_a$ with good approximation. All the other nuisances are then called ``ill-behaved''. Testing if the model is linear in a given nuisance only requires 3 calls to the model function and can be performed once for all provided the confidence contour is not too large. \smallbreak
The profiling over the well-behaved nuisances can be performed analytically using the Laplace approximation. Let $\eta_0$ be the mode of the nuisance distribution. If the nuisances are well-behaved, the log-likelihood around $\eta_0$ is quadratic with good approximation:
\begin{equation}
    \ell(p,\eta_0+\delta\eta)=\ell(p,\eta_0)+\sum_k\eval{\pdv{\ell}{\eta_k}}_{\eta_0}\delta\eta_k+\frac{1}{2}\sum_{k,\ell}\eval{\pdv{\ell}{\eta_k}{\eta_\ell}}_{\eta_0}\delta\eta_k\delta\eta_\ell.
\end{equation}
For clarity we define:
\begin{equation}
    [\vec g_\eta]_k=\eval{\pdv{\ell}{\eta_k}}_{\eta_0}\qc [\mathbf{H}_\eta]_{k\ell}=\eval{\pdv{\ell}{\eta_k}{\eta_\ell}}_{\eta_0},
\end{equation}
so that: 
\begin{equation}
    \ell(p,\eta_0+\delta\eta)=\ell(p,\eta_0)+\vec g_\eta\cdot\delta\vec\eta+\frac{1}{2}\delta\vec\eta\cdot\mathbf{H}_\eta\delta\vec\eta.
\end{equation}
At a local minimum $\eta^*$ of $\ell$, its gradient vanishes
\begin{equation}
    \grad\ell(p,\hat\eta)=\vec g_\eta+\mathbf{H}_\eta\delta{\vec\eta}^{\,*}=0\Longleftrightarrow \delta{\vec\eta}^{\,*}=-\mathbf{H}_\eta^{-1}\vec g_\eta,
\end{equation}
hence the analytic expression of the profiled likelihood reads:
\begin{equation}
    \hat\ell(p)=\ell(p,\eta_0+\delta{\vec\eta}^{\,*})=\ell(p,\eta_0)-\frac{1}{2}\vec g_\eta\cdot\mathbf{H}_\eta^{-1}\vec g_\eta.
\end{equation}
We still need to compute the Hessian of the likelihood function at the nuisances' central values. It reads explicitly:
\begin{equation}
    [\mathbf{H}_\eta]_{ij}=\sum_{k,\ell}\pdv{\ell_\mathcal{O}}{r_k}{r_\ell}\pdv{r_k}{\eta_i}\pdv{r_\ell}{\eta_j}+\sum_{k}\pdv{\ell_{\mathcal{O}}}{r_k}\pdv{r_k}{\eta_i}{\eta_j}+\pdv{\ell_\eta}{\eta_i}{\eta_j},
\end{equation}
and as $r_k$ is just the model's prediction shifted by a constant, 
\begin{equation}
    \pdv{r_k}{\eta_i}=\pdv{f_k}{\eta_i}.
\end{equation}
Near the minimum of $\ell$, the first derivatives of the likelihood function are close to zero, we can therefore neglect the second sum in the expression of the Hessian, leaving the so-called \textit{Gauss-Newton} matrix
\begin{equation}
    \mathbf{H}_\eta=J_\eta^TW_{\mathcal{O}}J_\eta+W_\eta,
\end{equation}
where $J_\eta$ is the Jacobian matrix of the model function $f$ w.r.t. the nuisance parameters and $W_{\mathcal{O}}$, $W_\eta$ are the local curvatures of the likelihood functions, which can be computed analytically using our copula distributions. The only numerically challenging task is to compute the first-order derivatives of the likelihood function, $\vec g_\eta$ and the Jacobian of the model function $J_\eta$. \smallbreak
In terms of the copula log-density $c(u)$ and the marginal densities $p_k(z)$, the curvature matrices read:
\begin{equation}
    W_{ij}=-\pdv{c}{u_i}{u_j}p_ip_j+\delta_{ij}\qty[-\pdv{c}{u_i}p'_i-\frac{p''_ip_i-{p_i'}^2}{p_i^2}],
\end{equation}
and the expressions of the first and second derivatives of the copulas' log-densities are analytical for the Gaussian and Student copulas. They are implemented explicitly in the statistical backend; the resulting expressions are lengthy but follow directly from differentiating the corresponding log-density with respect to the transformed normal or Student variables.

\newpage 

\bibliographystyle{elsarticle-num}
\bibliography{references}

\end{document}